\newcount\who
\font\tenmit=cmmi10
\font\tenrm=cmr10

\font\tit=cmbx10 scaled\magstep 3
\font\bigbf=cmbx10 scaled\magstep2
\def\centra#1{\vbox{\null\vskip 0.0 truecm\rightskip=0pt
	     plus1fill\leftskip=0pt plus1fill#1}}
\def\titolo#1{\baselineskip=22truept\parindent=0pt\centra{\tit #1}}

\def\sezione#1#2{\who=1\bigskip\medskip\bigskip
{\bigbf\noindent\hbox{#1.}
\hskip 0.5truecm #2}
\bigskip\medskip
\who=0\hfill}

\def\subsez#1#2{\ifnum\who=0\bigskip\goodbreak\else\medskip\fi
{\bigbf\noindent\hskip 0.8truecm\hbox{#1.}
\hskip 0.5truecm #2}\nobreak \ifnum\who=0\bigskip\fi\nobreak}

\def \m {\mu}
\def \ti {\tilde}
\def \G {\Gamma}
\def\ltord{\hbox{$\;\raise.4ex\hbox{$<$}\kern-.75em\lower.7ex\hbox{$\sim$}
		    \;$}}
\def\gtord{\hbox{$\;\raise.4ex\hbox{$>$}\kern-.75em\lower.7ex\hbox{$\sim$}
		    \;$}}
\def\subR{\hbox{\lower.2ex\hbox{$_R$}}}
\def\subh{\hbox{\lower.4ex\hbox{$_h$}}}
\def\subq{\hbox{\lower.2ex\hbox{$_q$}}}
\def\subrhat{\hbox{\lower.2ex\hbox{$_{{\bf \hat {\it r}}}$}}}
\def\sub0hat{\hbox{\lower.2ex\hbox{$_{{\bf \hat {\it 0}}}$}}}
\documentstyle[11pt]{article}

\begin{document}
\pagestyle{empty}
%\draft
%\twocolumn[\hsize\textwidth\columnwidth\hsize\csname
%@twocolumnfalse\endcsname

%\title{Relativistic radiation hydrodynamics
%and evaporation of cosmological quark drops}
\null\vskip -1.0 truecm
\titolo{Evaporation of cosmological quark drops
and relativistic radiative transfer }
\bigskip
\medskip
\centerline{ \bf{Luciano Rezzolla}} \par
\medskip
\centerline{{\it Scuola Internazionale di Studi
Avanzati, Trieste, Italy} }\par
\bigskip
\centerline{ \bf{John C. Miller}} \par
\medskip
\centerline{{\it Scuola Internazionale di Studi
Avanzati, Trieste, Italy} }\par
\centerline{{\it Department of Physics, University of
Oxford, England } }\par
\centerline{{\it Osservatorio Astronomico di Trieste,
Trieste, Italy } } \par
\bigskip
\begin{abstract}
\medskip
\noindent
We discuss the results of a full relativistic treatment of the
hydrodynamics of disconnected quark regions during the final stages of
the cosmological quark-hadron transition. In this study, which
represents a further development of a previous analysis of the
evaporation of cosmological quark drops, the effects of long range
energy and momentum transfer via electromagnetically interacting
particles are consistently taken into account. For this purpose, a set
of relativistic Lagrangian equations describing the evolution of the
strongly interacting fluids is coupled to a system of equivalent
equations accounting for the hydrodynamics of the fluid of
electromagnetically interacting particles. The complete set of
equations has then been solved numerically and results are presented
from this. The inclusion of relativistic radiative transfer produces
significantly different results, with the formation of high density
regions at the end of the drop evaporation being particularly
relevant. A comparison is made with results obtained for the previous
radiation-free model and the cosmological implications concerning
baryon number concentrations are briefly discussed.
\end{abstract}
\bigskip {PACS number(s): 47.55.Dz, 47.75.+f, 64.60-i, 98.80.Cq}\par
\bigskip
\centerline { SISSA Ref. 116/95/A (Oct 95)}
\vfill\eject
% ]
%\footnotetext{$^*$Electronic address:
%{\tt rezzolla@tsmi19.sissa.it} }\par
%\footnotetext{$^{\dag}$Electronic address:
%{\tt miller@tsmi19.sissa.it} }\par
%\medskip
%\vfill\eject

%-.-.-.-.-.-
%\null
\setcounter{page}{1}
\pagestyle{plain}
\vsize=20.0truecm
\hsize=14.0truecm
\baselineskip=14pt plus 2pt minus 1pt
\parindent=1.0truecm
\hoffset=0.0truecm
%-.-.-.-.-.-

\sezione{I}{Introduction}

	A phase transition at which the cosmological plasma of free quarks
and gluons was transformed into a plasma of light hadrons, is thought to
have occurred early in the history of the Universe. The physical
conditions for this transition to take place, date it back to a few
microseconds after the Big-Bang, when the Universe had a mean density of
the same order as nuclear matter ($\rho \sim 10^{15}$ g cm$^{-3}$), and a
temperature of the order of $100-200$ MeV. The quark-hadron transition
marks the end of the exotic physics of the very early Universe and the
beginning of the era of processes and phenomena which have a direct
counterpart in the high energy experiments now being carried out with
modern accelerators. It is also the last of the early Universe phase
transitions (at least within the standard picture) and so could be
relevant both as a potential filter for the relics produced by previous
transitions and also as a ``best candidate'' for the production of
inhomogeneities which could have survived to later epochs.

	It has not proved possible to determine the order of the
transition directly from QCD but, rather, its determination depends on
heavy lattice gauge calculations which rely on a number of simplifying
assumptions and uncertain parameters. Because of this, any consistent
modelling of the transition is immediately confronted by a major
uncertainty concerning the order of the transition. It is relevant to note
that while a continuous cosmological quark-hadron phase transition
\cite{kl94} would strongly prevent any dynamical production of primordial
inhomogeneities \cite{cs82}, the occurrence of these seems to be a rather
natural consequence of a first order transition. We here follow this
latter scenario and investigate the hydrodynamical mechanisms which could
lead to the production of inhomogeneities at the end of a first order
cosmological quark-hadron transition. We note that this picture is, in
fact, favoured by recent lattice computations which include the effects of
two degenerate light $u$ and $d$ quarks and a heavier $s$ quark (of up to
400 MeV) \cite{ikkry94,ikksy94,k95,ikksy95}, and clearly indicate the
existence of a double state signal for the quark gluon plasma.

	The aim of the present work is to discuss the final stages of the
transition, which we define to be those when most of the strongly
interacting matter in the Universe is already in the form of light
hadrons. The temperature jump between the quark and hadron phases is then
no longer extremely small and the rate at which the quark-gluon plasma is
transformed is no longer controlled by the overall expansion of the
Universe. The quark regions have become disconnected, with a mean
separation comparable with the distance between bubble nucleation sites,
and tend to assume a spherical shape under the effects of surface tension.
A new dynamical time scale for the evolution of the transition then
enters and this is directly related to the rate at which the quark drops
shrink by losing material ({\it i.e.} by ``evaporating'').

	In a recent paper \cite{rmp95}, we have investigated and discussed
the relativistic hydrodynamics of an isolated quark drop during the final
stages of the transition in the simplified picture where long range energy
and momentum transfer is not included. In particular, we demonstrated the
existence of a self-similar solution for the hydrodynamics of an isolated
contracting spherical system and showed how this is, in fact, attained by
an evaporating quark drop. A most important feature turned out to be the
possibility of maintaining a constant quark phase compression during most
of the stages of the drop evaporation. This had the consequence of
preventing any large increase in the final value of the compression factor
in the quark phase (the maximum relative increase computed was of the
order of $40\%$), thus limiting the possibility of producing significant
peaks of baryon number density as the drop disappears. In that work we
considered the fluids of strongly interacting particles and that of
electromagnetically interacting particles as always being coupled,
neglecting the effective decoupling which must occur when the drop
dimensions are comparable with the mean free path of the radiation
particles.

	In the present paper we extend the earlier study to the case where
the effects of long range energy and momentum transfer are not neglected.
To do this, a problem of relativistic radiative transfer needs to be
solved at around the time of the decoupling between the two fluids, and an
extended set of equations has to be solved numerically. In doing this we
make use of the experience gained in the study of the related problem of
long range energy and momentum transfer during the growth of a hadron
bubble \cite{rm94, mr95}, and also use the mathematical and numerical
apparatus developed there.

	The following is a summary of the organization of the paper. In
Section II we review the essential features of the PSTF (Projected
Symmetric Trace Free) tensor formalism \cite{t81} adopted for the solution
of the relativistic radiative transfer; in Section III we introduce the
set of hydrodynamical equations for the standard fluids and discuss how
these couple to the equivalent ones for the radiation fluid. Section IV
contains a discussion of the solution of the equations at the interface,
where junction conditions and characteristic equations are solved. In
Section V we briefly review the method for the numerical computation and
the choice for the initial conditions. Numerical results are presented in
Section VI and Section VII contains a brief discussion of their
cosmological implications regarding the formation of baryon number
inhomogeneities, which may be important in connection with primordial
nucleosynthesis \cite{afm87,jfmk94} and also with the production of
primordial magnetic fields \cite{co94}. Finally, conclusions are presented
in Section VIII. We adopt units for which $c = \hbar = k_{_B} = 1$; Greek
indices are taken to run from $0$ to $3$ and partial derivatives are
denoted with a comma.

\sezione{II}{Relativistic hydrodynamic equations for the \hfill
\ \hfill\break  \vskip 0.01 truecm
\vskip -0.3 truecm\hskip 0.5truecm radiation fluid\hfill \ \hfill}

	In the present Section and in the following one, we discuss the
formulation of the system of equations which we use for studying the
hydrodynamics of a contracting spherical drop, with radiative exchange of
energy and momentum being included. We consider a ``two-fluid''
hydrodynamical model, and refer to the fluid composed of strongly
interacting particles as the ``standard fluid'', and to the one composed
of electromagnetically interacting particles (mainly photons, electrons,
muons and their antiparticles) as the ``radiation fluid''. While the
particles of the first fluid have a typical interaction scale length of
the order of 1 fm, the particles of the second fluid have a larger
interaction scale length, probably between $5 \times 10^3$ and $10^4$ fm.
Neutrinos, which have a much larger interaction scale length (of the order
of $10^{13}$ fm $=1$ cm), provide no effective contribution at the scale
which we are mainly concerned with here and will in general be neglected.

	Because of the difference between the scales of the internal
interactions in the two fluids, it is possible to consider two asymptotic
regimes during drop evaporation. When the drop has a radius $R_s$ which is
much larger than the mean free path (m.f.p.) of the radiation fluid
particles $\lambda$, ($R_s \gg \lambda$), it is reasonable to consider the
two fluids as effectively coupled and behaving dynamically as a single
fluid within each phase. In this case the treatment is simplified and the
contribution of the different particle species can be taken into account
by a suitable specification of the number of degrees of freedom in the
equations of state for the two phases. On the other hand, when the drop
has dimensions which are much smaller than the radiation m.f.p., ($R_s \ll
\lambda$), the opposite asymptotic regime is reached, where the drop has
become effectively transparent to the radiation fluid particles. In this
case, the decoupling between the two fluids can be taken into account by
eliminating the number of degrees of freedom of the radiation fluid
particles in the equations of state.

	During drop evaporation there will be a stage (at $R_s \approx
\lambda$) when the two fluids will start to decouple and long range
exchange of energy and momentum will act in the direction of smoothing
the discontinuities produced by the temperature difference between the
two phases. In order to follow the effects of this transient process,
it is necessary to adopt a treatment in which the radiative transfer
problem and the hydrodynamical problem are solved simultaneously. For
this purpose we have implemented the mathematical apparatus developed
for studying the related problem of the progressive coupling between
the standard fluid and the radiation fluid during the growth of a
hadron bubble \cite{rm94, mr95} and the reader is referred to this
work for details of the derivation of the equations. We here limit
ourselves to outlining the assumptions and main results of the PSTF
tensor formalism which has been adopted.

	As standard in radiation hydrodynamics, we need to solve the
radiative transfer equation, which relates the properties of the
radiation field (described by the photon distribution function) to the
sources and sinks of the field and to the dynamics of the underlying
medium when this is not stationary. For doing this, we make use of the
frequency integrated PSTF tensor formalism in which the relativistic
generalization of the radiative transfer equation is transformed into
an infinite hierarchy of partial differential equations involving an
infinite number of moments of the radiation intensity and of the field
sources (the latter are referred to as source functions). A
particularly attractive feature of the PSTF tensors, which are
suitably defined at each point in the projected tangent space to the
fluid four-velocity, is that they become effectively scalars when a
global planar or spherical symmetry is present. Because of this, the
PSTF formalism is particularly suitable for solving the radiative
transfer problem for a contracting quark drop, since in this case,
spherical symmetry enters as a natural consequence of the drop
dynamics.

	As in any infinite series expansion strategy, all of the
properties of the radiation field are known exactly only when the
infinite hierarchy of moments is determined. However, this is never
possible in practice and a truncation at a finite order in the moment
expansion is therefore necessary. This has two main consequences:
firstly it introduces an overall intrinsic approximation in the
determination of the radiation variables, and secondly it requires the
introduction of a closure relation which specifies the value of the
highest moment used in terms of lower ones. This supplementary
equation, which should be derived on the basis of physical
considerations, is somewhat heuristic and the form used for it is
typically related to the specific problem under investigation.

	As in \cite{rm94, mr95}, we here truncate the infinite hierarchy
of moments at the second order, thus making use of the first three scalar
moments $w_0$, $w_1$ and $w_2$ and of the first two source functions $s_0$
and $s_1$. A truncation at the second order, which introduces an intrinsic
overall error of the order of 15$\%$, has a number of interesting and
convenient features. Firstly, all of the scalar moments retained have
direct physical interpretation, with $w_0$ and $w_1$ being the energy
density and flux of the radiation in the rest frame of the standard fluid,
and with $w_2$ representing the shear stress scalar of the radiation.
Secondly, the moments used are only those appearing explicitly in the
stress-energy tensor for the radiation fluid $T_{_R}^{\alpha \beta}$,
which, at any order, has the form
\vfill\eject

\begin{equation}
\label{rest}
T^{\alpha \beta}_{_R} = {\cal M} u^{\alpha} u^{\beta} +
{\cal M}^{\alpha}u^{\beta}
+ {\cal M}^{\beta}u^{\alpha}  + {\cal M}^{\alpha \beta} +
{1\over 3} {\cal M} P^{\alpha \beta} ,
\end{equation}
where $P^{\alpha \beta}$ is the projection operator orthogonal to the
fluid four-velocity $u^{\alpha}$, and the first three PSTF moments
${\cal M}$, ${\cal M}^{\alpha}$, ${\cal M}^{\alpha \beta}$, are
related to the equivalent scalar moments via the expressions

\begin{equation}
{\cal M} = w_0 ,
\end{equation}

\begin{equation}
{\cal M}^{\alpha}=w_1 e^{\alpha}_{{\bf {\hat {\it r}}}} ,
\end{equation}

\begin{equation}
{\cal M}^{\alpha\beta}=
w_2 \biggl({e^{\alpha}_{{\bf \hat{\it r}}}
e^{\beta}_{{\bf \hat {\it r}}} -
{1\over 2} e^{\alpha}_{{\bf \hat\theta}}
e^{\beta}_{{\bf \hat \theta}} -
{1\over 2} e^{\alpha}_{{\bf \hat\varphi}}
e^{\beta}_{{\bf \hat \varphi}}} \biggr) .
\end{equation}

\noindent Here $({\bf e}\sub0hat,\; {\bf e}\subrhat,\; {\bf e}_{{\bf \hat
\theta}},\; {\bf e}_{{\bf \hat \varphi}})$ is the orthonormal tetrad
carried by an observer comoving with the standard fluid. A particular
advantage of truncating at the second order is that it is then possible to
avoid the use of iterative methods for the derivation of the equations
governing the hydrodynamics of the radiation fluid, which can instead be
derived by means of the standard conservation laws of energy and momentum
for the radiation fluid. We adopt Lagrangian coordinates comoving with the
standard fluid and having their origin at the centre of the drop and write
the (spherically symmetric) line element as

\begin{equation}
\label{le}
ds^2 = -a^2 dt^2 + b^2 d\mu^2 +
R^2 ( d\theta^2 + {\rm sin}^2 \theta \ d\varphi^2) ,
\end{equation}
where $\mu$ is a comoving radial coordinate and $R$ is the associated
Eulerean coordinate. The PSTF equations can then be written as

\begin{equation}
\label{ren}
-u_{\alpha} T^{\alpha \beta}_{{_R}\;\; ;\beta} = s_0 ,
\end{equation}

\begin{equation}
\label{rmom}
P_{\mu \alpha} \  T^{\alpha \beta}_{{_R}\;\; ;\beta} = {s_1 \over {b}} ,
\end{equation}

\begin{equation}
\label{clrl}
w_2 = f_{\!_E} w_0.
\end{equation}

	Equation (\ref{clrl}) is the closure relation and specifies a
connection between the second and the zeroth moments in terms of a
variable Eddington factor $f_{\!_E}$, which is an indicator of the
degree of anisotropy of the radiation. This Eddington factor can take
values ranging from $0$ for complete isotropy (which, for example, is
reached when the radiation fluid and the standard fluids are totally
coupled) to $2/3$ for complete anisotropy (which could, in principle
be reached when the two fluids are effectively decoupled). An
expression for $f_{\!_E}$ has to be supplied and in doing this it is
important that the correct asymptotic behaviour in any relevant limits
is preserved and that the form chosen provides a suitably smooth join
between the physical limits. Experience has shown \cite{ntz91, ntz93,
mr95} that as long as these requirements are met, results do not
usually depend sensitively on the precise form used for $f_{\!_E}$. As
in \cite{rm94, mr95}, we have here used for the Eddington factor the
expression

\begin{equation}
\label{fe}
f_{\!_E} \equiv {8 u^2 / 9 \over {(1 + 4 u^2/3)}}\left({\lambda \over
{\lambda + R}}\right) ,
\end{equation}
which is the product of an exact expression accounting for the Doppler
effects of motion with respect to a uniform radiation field, together
with a corrective term (the one in the large parentheses) which
provides the required physical join between the optically thin and
optically thick limits. The scalar source functions $s_0$ and $s_1$
appearing in (\ref{ren}), (\ref{rmom}), represent the sources or sinks
of energy and momentum between the two fluids, and are expressed as

\begin{equation}
\label{s0}
s_0 = {1 \over {\lambda}} (\epsilon - w_0) + \left( s_0 \right)_{sc},
\end{equation}

\begin{equation}
\label{s1}
s_1 = - {w_1 \over {\lambda}} ,
\end{equation}
where $\epsilon$ is the energy density for radiation in thermal
equilibrium with the standard fluid ({\it i.e.} it is the equivalent
of a local emissivity), and the term $(s_0)_{sc}$ expresses the
contribution to the energy source due to non-conservative scatterings.
Assuming a \hbox{black-body} expression for $\epsilon$, we have

\begin{equation}
\label{eps}
\epsilon = g_r \left({\pi^2 \over{30}}\right) T_{_F}^4 ,
\end{equation}

\noindent
with $g_r$ being the number of degrees of freedom of the radiation fluid
and $T_{_F}$ the local temperature of the standard fluid. Obtaining a
suitable expression for $(s_0)_{sc}$ (which in general depends on the
details on the problem under investigation) is particularly problematic in
the present case where the number and the complexity of all the possible
particle interactions prevent us from having an exact and simple expression.
For this reason, we have adopted a phenomenological view and have
expressed $(s_0)_{sc}$ in terms of the simple absorption and emission
factor

\begin{equation}
\label{s0c}
(s_0)_{sc} ={\alpha_2 \over {\lambda}} (\epsilon - w_0),
\end{equation}

\noindent
where $\alpha_2$ is an adjustable coefficient ranging between zero and
one. Within a cosmological context it is reasonable to assume
$\alpha_2 \approx 1$; a discussion of the differences caused by
varying $\alpha_2$ will be presented in Section VI.

	Expressing equations (\ref{ren}) and (\ref{rmom}) explicitly
in terms of our metric, we obtain the following two equations of
relativistic radiation hydrodynamics

\begin{equation}
\label{w0}
(w_0)_{,\> t}+{a\over b}(w_1)_{,\>\mu}
+{4\over 3}\biggl({{b_{,\> t}\over b}+{2R_{,\> t}\over R}}\biggr)w_0
+{2a\over b}\biggl({{a_{,\>\mu}\over a}+{R_{,\>\mu}\over R}}\biggr)w_1
+\biggl({{b_{,\> t}\over b}-{R_{,\> t}\over R}}\biggr)w_2=as_0 ,
\end{equation}

\begin{equation}
\label{w1}
(w_1)_{,\> t}+{a\over b}\biggl({{1\over 3}w_0+w_2}\biggr)_{\!\! ,\mu} +
{4a_{,\>\mu}\over {3b}}w_0
+2\biggl({{b_{,\> t}\over b}+{R_{,\> t}\over R}}\biggr)w_1
+{a\over b}\biggl({ {a_{,\>\mu}\over a}+{3R_{,\>\mu}\over R}}\biggr)w_2
=as_1 ,
\end{equation}

\noindent which, together with equation (\ref{clrl}) provide a consistent
description of the transfer of energy and momentum via the radiation
fluid. During the final stages of the evaporation ({\it i.e.} for $R_s \ll
\lambda$), the drop medium is locally optically thin and the energy
density of the radiation fluid becomes uniform in the Eulerean frame.
Under these circumstances (which are similar to those encountered in the
early stages of bubble growth \cite{mr95}), there is a tendency for
numerical instabilities to appear, related to near cancellation problems
and to almost diverging expressions in the characteristic form of
equations (\ref{w0}) and (\ref{w1}). Experience has shown that a
satisfactory numerical solution of the above equations is then possible
only if they are rewritten in terms of new variables, defined by the
following transformations

\begin{eqnarray}
\label{tilde1}
{\ti w_0} = w_0 - (w_0)^{*} = w_0 -
\left({1+{4\over 3} u^2 }\right)(w_0)_{_N},  \\
\label{tilde2}
{\ti w_1} = w_1 - (w_1)^{*} = w_1+{4\over 3} u\Gamma (w_0)_{_N}, \\
\label{tilde3}
{\ti w_2} = w_2 - (w_2)^{*} = w_2 - {8\over 9}u^2(w_0)_{_N} ,
\end{eqnarray}
\noindent
where

\begin{equation}
\label{rt}
u={1\over a}R_{,\> t} \;,
\end{equation}

\begin{equation} \label{gamma} \Gamma={1\over
b}R_{,\>\mu}=\biggl({1+u^2-{2GM\over R}}\biggr)^{\!1/2} . \end{equation}
\noindent Here $u$ is the radial component of fluid four velocity in the
associated Schwarzschild (Eulerian) frame, $\Gamma$ is the general
relativistic analogue of the Lorentz factor and $M$ a generalized mass
function. The new ``tilde'' variables are expressed as differences between
the standard moments and the expressions for these moments resulting from
considering the motion of the fluid relative to a uniform radiation field
having energy density $(w_0)_{_N}$ in the Eulerean frame. This will here
coincide with the initial value of the radiation energy density in the
hadron phase, which we have therefore taken as the reference value. Making
use of (\ref{tilde1})--(\ref{tilde3}), it is then possible to obtain the
following transformed radiation hydrodynamics equations \cite{mr95}

\begin{eqnarray}
\label{w0t}
({\ti w_0})_{,\> t} +
a{\ti w_0} \left [{1 \over {R^2}}
\left ( {4\over 3}+ f_{\!_E} \right )
(uR^2)_{,\> R} - {3 u f_{\!_E} \over R} \right ]+
{\G \over {a R^2}} ({\ti w_1} a^2 R^2)_{,\> R} \hfill \nonumber \\
+ a {4 \over {3 R}} ({ w_0})_{_N}
\left [ f_{\!_E} ( {3\over 4}  + u^2) - {2 \over 3} u^2 \right ]
\left [{1 \over R} (u R^2)_{,\> R} - 3 u \right]  + a s_0 \nonumber \\
 - {4 \over 3 } a ({ w_0})_{_N} G \left [ 4 \pi u R
\left ( 2p - e - {{w_0} \over 3} + 2 {w_2} -
{u \over {\G }} {w_1} \right ) - {M \over R }
\left ( 2 u_{,\> R} + {u \over R}\right ) \right ]  \hfill \nonumber \\
\hfill{} - { 4 \pi a G R \over {\G}} \left ( {4 \over 3} {w_0} +
{w_2} \right )  w_1 =0 \ , \hskip 2.0 truecm
\end{eqnarray}

\begin{eqnarray}
\label{w1t}
({\ti w_1})_{,\> t} + 2 {\ti w_1} {a \over R} (u R)_{,\> R} +
a \G \left ({{\ti w_0} \over 3} + {\ti w_2} \right)_{,\> R}+
\G \left( {4 \over 3}{\ti w_0} + {\ti w_2}\right) a_{,\> R}
+ {3 a \G {\ti w_2}\over R} \hfill \nonumber \\
+ a s_1+ {4 \over 3 } a ({w_0})_{_N} \G G \left [ 4 \pi R
\left ( p + {{ w_0} \over 3} + { w_2} -
{u \over {\G }}{\ w_1}\right ) +
{M \over {a^2 R^2}}(a^2 R)_{,\> R} \right ] \hfill \nonumber \\
- {8 \pi a G R {w_1}^2 \over {\G}}=0 \ . \hskip 4.0truecm
\end{eqnarray}
\noindent
where $\rho$ is the {\it relative compression factor}, and $e$ and $p$
are the energy density and the pressure of the standard fluids. Note
that it has been convenient here to replace the the partial derivatives
with respect to $\m$ by the equivalent derivatives with respect to $R$
and that, for compactness, the radiation variables which are
multiplied by the gravitational constant $G$ are not transformed
according to (\ref{tilde1})--(\ref{tilde3}). Equations (\ref{w0t}),
(\ref{w1t}) represent our final form of the hydrodynamical equations
for the radiation fluid and need to be solved together with the
corresponding hydrodynamical equations for the combined fluids, which
will be discussed in the next Section.

\vfill\eject
\sezione{III}{Relativistic hydrodynamic equations for the \hfill\break
\vskip 0.01 truecm
\vskip -0.3 truecm\hskip 0.7truecm standard fluid\hfill}

	The formal derivation of the hydrodynamical equations for the
standard fluids is more standard. For this purpose in fact, it is possible
to make use of the ordinary conservation equations for energy and momentum
of the combined fluids ({\it i.e.} standard fluid plus radiation fluid)
together with the continuity equation for the standard fluid. The
equations which are then obtained, can be rewritten in a more familiar
form by combining them with the Einstein field equations expressed in
terms of the metric (\ref{le}) and of a ``total'' stress-energy tensor
(the sum of the one for the radiation fluid and of the one for the
standard fluid). The set of equations is then \cite{rm94}

\begin{equation}
\label{u}
u_{,\> t}=-a\biggl[{{\Gamma\over b}\biggl({p_{,\>\mu}+
bs_1\over {e+p}}\biggr)
+ 4\pi G R \biggl({p+{1\over 3}w_0 + w_2}\biggr)
+ {G M\over {R^2}}}\biggr] ,
\end{equation}

\begin{equation}
{(\rho R^2)_{,\> t}\over {\rho R^2}}=
-a\biggl({{u_{,\>\mu}-4\pi G b R w_1\over {R_{,\>\mu}}}}\biggr) ,
\end{equation}

\begin{equation}
e_{,\> t}=w\rho_{,\> t}-as_0 ,
\end{equation}

\begin{equation}
{(aw)_{,\>\mu}\over {aw}}=
-{w\rho_{,\>\mu}-e_{,\>\mu}+bs_1\over {\rho w}} ,
\end{equation}

\begin{equation}
M_{,\>\mu}=4\pi R^2 R_{,\>\mu}\biggl({e + w_0 +
{u\over {\Gamma}}w_1}\biggr) ,
\end{equation}

\begin{equation}
\label{b}
b = {1 \over { 4 \pi R^2 \rho }},
\end{equation}

\begin{equation}
\label{w}
w = {(e+p) \over {\rho}}\;,
\end{equation}
\noindent
where $a$ and $b$ are the metric coefficients and $w$ is the specific
enthalpy of the standard fluids. The compression factor $\rho$
expresses the variation in the proper volume of comoving elements of
the standard fluid and for a classical standard fluid it can be
replaced by the rest mass density. The set of equations
(\ref{rt})--(\ref{gamma}) and (\ref{u})--(\ref{w}) needs to be
supplemented with equations of state for both phases of the strongly
interacting matter. For small net baryon number and taking the
hadronic medium to consist of massless point-like pions, it is
appropriate to describe the hadron plasma as an ultra-relativistic
fluid, for which

\begin{equation}
\label{eosh}
e\subh = (g\subh + g_r) \left({\pi^2 \over {30}}\right) T^4_h \ ,
\hskip 4.0truecm p\subh = {1 \over 3} e\subh \ ,
\end{equation}
while the quark phase can be effectively described by the {\it Bag model}
equation of state

\begin{equation}
\label{eosq}
e\subq = (g\subq + g_r)
\left({\pi^2 \over {30}}\right) T^4_q + B \ ,
\hskip 1.0truecm
p\subq = (g\subq + g_r)
\left({\pi^2 \over {90}}\right) T^4_q - B \ ,
\end{equation}

\noindent
where $B=(\pi^2/90)(g\subq-g\subh)T^4_c$ is the ``bag''
constant, $T_c$ is the critical temperature for the transition and
$g\subq$, $g\subh$, $g_r$ represent the number of effective degrees of
freedom of the quark matter, the hadronic matter and the radiation
particles respectively. Note that equations (\ref{eosh})--(\ref{eosq}),
(with $g\subq=37$, $g\subh=3$, and $g_r=9)$, apply when one considers the
standard fluids and the radiation fluid as totally coupled, which is the
case when the drop has dimensions $R_s \gg \lambda$. However, at the
decoupling, and for all the subsequent stages of the drop evaporation,
they need to be corrected by removing the additional number of degrees of
freedom of the radiation fluid particles.

	Equations (\ref{u})--(\ref{eosq}), together with equations
(\ref{rt})--(\ref{w1t}) and (\ref{clrl}) represent the full system of
hydrodynamical equations in the presence of long range energy and
momentum transfer via electromagnetically interacting particles.
However, the numerical solution of these equations must necessarily
take into account the presence of a discontinuity between the phases
of the strongly interacting matter at the drop surface. For this
reason an appropriate treatment needs to be made of the conditions at the
phase interface, and the way this has been accomplished is illustrated
in the next Section.

\sezione{IV}{Solution at the interface}

	The presence of an interface dividing the two phases of the
strongly interacting matter introduces a number of complications when
the set of the hydrodynamical equations discussed in the previous
Sections is to be solved numerically. A first complication is related
to the nature of the drop surface and to the way in which it should be
described. Given the lack of detailed knowledge of the microphysics
within the phase interface and the fact that the width of the
interface is generally small compared with the typical radial scale of
the problem, it is convenient to treat it as a discontinuity surface
across which rapid variations in the fluid variables occur. It is then
possible to join the solution of the hydrodynamical equations on
either side by imposing a number of relativistic junction
conditions. For the particular case of a discontinuity surface with
associated physical properties, (a non-vanishing stress energy tensor
and an intrinsic curvature), the junction conditions can best be
derived using the Gauss--Codazzi formalism \cite{m86,mp89,mr95}.
Assuming that the surface tension $\sigma$ is independent of
temperature, (\hbox{$\sigma = \sigma_0 T_c^3$, with $0 \le \sigma_0
\le 1$}), the conservation of energy and momentum across the interface
are expressed as

\begin{equation}
\label{ejc}
[(e+p)ab]^{\pm}=0 ,
\end{equation}

\begin{equation}
\label{mjc}
[eb^2{\dot \mu_{_S}}^2+pa^2]^{\pm}=-{\sigma f^2\over 2}\biggl\{
{{1\over {ab}}{d\over {dt}}\biggl({b^2\dot \mu_{_S}\over f}\biggr) +
{f_{,\>\mu}\over {ab}} +
{2\over {fR}}{(b\dot \mu_{_S} u + a\Gamma)}}\biggr\}^{\pm} ,
\end{equation}

\noindent
where $[A]^{\pm} = A^+ -  A^-$,\  $\{A\}^{\pm} = A^+ + A^-$,\
$\mu_{_S}$ is the interface location, $\dot \mu_{_S} = d\mu_{_S}/dt$,
$f = (a^2 - b^2 \dot \mu^2_{_S} )^{1/2}$ and the superscripts $^{\pm}$
indicate quantities immediately ahead of and behind the interface.
Note that the energy density and the pressure appearing in
(\ref{ejc}), (\ref{mjc}) are the sum of those for the standard fluids
and for the radiation fluid when these are totally coupled. At the
decoupling however, and for all the following stages of the drop
evaporation, it is necessary to supplement the equations (\ref{ejc}),
(\ref{mjc}) (which will then refer to the standard fluids only), with
the equivalent junction conditions for energy and momentum of the
radiation fluid. Assuming that there is no interaction of the
radiation fluid with the matter in the phase interface, the energy and
momentum junction conditions for the radiation are then just the
continuity conditions

\begin{eqnarray}
\label{erjc}
\biggl[ ab {\dot \mu_{_S}} \left( {4\over 3}+f_{\!_E} \right)
{\tilde w_0} -(a^2+b^2 {\dot \mu_{_S}}^2) {\tilde w_1}+
({w_0})_{_N} \biggl\{ a b {\dot \mu_{_S}}
\left( 1 + {4 \over3 } u^2 \right)
\left( {4 \over 3} + f_{\!_E} \right) \nonumber \\
 + {4 \over 3} u \Gamma (a^2+b^2 {\dot \mu_{_S}}^2)
\biggr\}\biggr]^{\pm}=0 \ , \hskip 1.5 truecm
\end{eqnarray}

\begin{eqnarray}
\label{mrjc}
\biggl[\left\{a^2\left({1\over 3} + f_{\!_E}\right)+
b^2{\dot \mu_{_S}}^2\right\}{\tilde w_0}
-2ab\dot \mu_{_S}{\tilde w_1} \hskip 5.0 truecm \nonumber \\
+ ({w_0})_{_N} \left\{\left( 1 + {4 \over3 }u^2\right)
\left[ a^2\left({1\over 3} + f_{\!_E}\right)+
b^2{\dot \mu_{_S}}^2 \right]+
{8\over 3}abu\Gamma \dot \mu_{_S} \right\} \biggr]^{\pm}=0 .
\end{eqnarray}
\noindent
Other supplementary junction conditions follow from continuity
across the interface of the metric quantities $R$, $dR/dt$, and $ds$
\vfill\eject

\begin{equation}
[R]^{\pm}=0 ,
\end{equation}

\begin{equation}
[au+b \dot \mu_{_S} \Gamma]^{\pm}=0 ,
\end{equation}

\begin{equation}
[a^2-b^2 {\dot \mu_{_S}}^2]^{\pm}=0 ,
\end{equation}

\noindent
and from the time evolution of the mass function $M$

\begin{eqnarray}
\label{M}
{d \over {dt}}[M]^{\pm} = 4\pi R^2_{_S}
\biggl[  b \Gamma {\dot \mu_{_S}}
\left\{ \left( e + w_0 + {u \over \Gamma}w_1 \right) \right\} -
\hskip 3.0truecm \nonumber \\
a u \left\{ { p + \left( { {1\over 3} + f_{\!_E} } \right)
w_0 + {\Gamma \over u} w_1 }
\right\}  \biggr]^{\pm} ,
\end{eqnarray}

\medskip
\noindent
where the initial jump in mass across the interface is taken to be
\hbox{$[M]^{\pm}=4\pi R^2_{_S}\sigma$}.

	The concept of the phase interface as a perfect discontinuity
surface needs careful interpretation in the context of a numerical
calculation and in this case it is important to bear in mind that the
interface should not be considered as strictly infinitesimal. In the
present situation, in which the computer code follows the drop evaporation
with an increasing spatial resolution through a number of orders of
magnitude in radius, the interface should be thought of as having an
effective width which is always smaller than the minimum length scale
resolvable on the grid. This means that the numerical code will treat as
discontinuous any change in the physical variables which cannot be
resolved on the grid. This feature is particularly relevant at the
decoupling, because at that stage the long range energy and momentum
transfer introduces features of the flow on length scales which were not
previously resolved when the standard fluids and the radiation fluid were
considered as coupled. When the decoupling is allowed to start, the
effective width of the phase interface is abruptly decreased to that
appropriate for the strongly interacting matter alone and, as a
consequence, changes across it which were previously discontinuous are
allowed to smooth down and assume the profiles produced by the radiative
transfer. In nature, the change between the two different situations is
progressive and regular, but when described on an finite grid, it occurs
discontinuously. Doing this requires particular care and in the next
Section we discuss the details of the computational strategy which has
been implemented in order to perform this change.

	A further complication regarding the solution at the interface
arises because of the dynamical properties of the drop surface treated as
a reaction front. General considerations about the nature of the
transition \cite{gkkm84, bp93, mp89} lead us to consider the transition as
taking place by means of a weak deflagration front ({\it i.e.} by means of
a discontinuity surface moving subsonically relative to the media both
ahead and behind) \cite{ll93,cf76}. Weak deflagrations are intrinsically
under-determined and require the specification of one additional condition
giving the rate at which the quark matter is transformed into hadrons at
the phase interface \cite{cf76}. A simple and satisfactory expression can
be obtained by setting the hydrodynamical flux $F_{_H}$ into the hadron
region equal to the net thermal flux $F_{_T}$ into it

\begin{eqnarray}
\label{flux}
F_{_H}=-{{aw\dot\mu_{_S}}
\over{4\pi R^2_{_S}(a^2-b^2\dot \mu_{_S}^2)}}
=\left({\alpha_1 \over 4}\right)
(g_h+g_r) \left({\pi^2 \over {30}}\right)
(T_q^4-T_h^4)=F_{_T},
\end{eqnarray}

\noindent
where $\alpha_1$ is an accommodation coefficient ($0\le \alpha_1 \le
1$) containing information about the ``transparency'' of the phase
interface to the thermal flux and is, at least in principle,
calculable from theory.

	It is very important to ensure that the correct causal structure
is preserved when following the motion of a weak deflagration front as a
discontinuity surface. A careful numerical investigation reveals that the
only satisfactory way of accomplishing this is by making use of a
characteristic method in which the system of partial differential
equations is rewritten as a system of ordinary differential equations
along specific curves in the space-time (the characteristic curves). The
correct causal connection is then preserved as the characteristic curves
are the world-lines along which information propagates through the media.
We here make use of the same system of characteristic equations employed
for the growth of a hadronic bubble and for compactness their lengthy
expressions will not be repeated here (we refer the reader to
\cite{rm94,mr95}, where the details of their mathematical derivation is
also given). Figure 1 shows the Lagrangian space-time configuration of the
characteristic curves adjacent to the interface for evolution of the
system from time level $t$ to a subsequent time level \hbox{$t + \Delta
t$}, with different line types distinguishing the different types of
fluid. Note that, the gradients of the corresponding characteristics on
the two sides of the phase interface can be different and that the
difference between the sound speeds in the radiation fluid \hbox{[$(1/3
+f_{\!_E})^{1/2}$]} and in the standard fluid ($c_s$), is greatly
magnified in the figure.

\bigskip
\bigskip
\bigskip
\centerline{\vbox{\hsize 12.0truecm\baselineskip=12pt\noindent\tenrm
\textfont1=\tenmit
Figure 1. The configuration of characteristic curves near the phase
interface drawn in the Lagrangian coordinate frame. }}
\bigskip
\bigskip
\bigskip

	As discussed above, the solution of the hydrodynamical
equations at the phase interface is rather complicated and requires
great care. However, once the set of radiation fluid and standard
fluid hydrodynamical equations is solved along the characteristic
curves (which are shown schematically in Figure 1) and the relevant
junction conditions are imposed across the phase interface, a
consistent numerical evolution of a weak deflagration front can then
be computed.

\sezione{V}{Numerical strategy and initial conditions}

	For following the evaporation of a quark-gluon drop and taking
into account the progressive exchange of energy and momentum which at a
certain stage will take place between the radiation fluid and the strongly
interacting fluids, we have here made use of the experience and numerical
codes developed for studying the related problems of radiative transfer
for a growing hadron bubble \cite{mr95} and of evaporation of a quark
drop in the absence of long range energy and momentum transfer
\cite{rmp95}. The result of this has produced a code which embodies the
main features of the previous ones and we will briefly describe this
here, referring the reader to the previous papers for further details.

	As with its predecessors, the present code makes use of a
composite numerical technique in which a standard Lagrangian
finite-difference method is used to solve the hydrodynamical equations in
the bulk of each phase, while a system of characteristic equations and a
set of junction conditions are solved in the regions adjacent to the phase
interface. The grid is Lagrangian and spherically symmetric with its
origin at the centre of the drop. In order to follow the solution over a
number of orders of magnitude in the spatial coordinate $\mu$, the grid
has variable spacing with the width of each successive zone being twice
that of the previous one ({\it i.e.}
\hbox{$\Delta\mu_{j+1/2}=2\;\Delta\mu_{j-1/2}$}), apart from the two
central zones which have equal width.

	The specification of the initial conditions for the system of
hydrodynamical equations has been guided by the existence of a
self-similar solution for an isolated contracting physical system
which was demonstrated in \cite{rmp95}. When there is no intrinsic
length scale influencing the problem (as in the case of an isolated
evaporating spherical drop for which surface tension is not yet playing a
significant role), it is possible to write the set of hydrodynamical
equations in terms of a single dimensionless independent variable and
to find a similarity solution. The time evolution of the system is
scale independent and reproduces itself at any instant. It is
important to stress that this is a general feature of the similarity
solution and holds for any dimensions of the system satisfying the
above assumptions.

	However, in the case of an evaporating quark drop during the
cosmological phase transition, we do not expect the self-similarity to
hold at all stages. Rather, it is necessary to establish an interval in
the drop dimensions within which such self-similar behaviour is expected
to take place. While the lower limit in the drop radius can easily be
estimated from the magnitude of the surface tension $\sigma$ associated
with the phase interface (the surface tension introduces a natural length
scale and the drop evaporation is no longer scale free when surface
effects become relevant), the definition of the upper limit is more
uncertain. In this case, it is necessary to determine an initial scale at
which the quark drops can be considered physically disconnected, so that
the distance between the centers of two neighbouring drops is larger than
the sum of their respective ``sonic radii'' (see \cite{rmp95} for a
definition). The value of this is not yet established and its
determination would require a detailed hydrodynamical study of the
intermediate stages of the transition, which we consider to be the ones
after the hadron bubbles have coalesced and the quark regions have started
to become disconnected. Simple geometrical considerations
suggest that the mean separation between quark regions at bubble
coalescence would be of the order of the mean separation of bubble
nucleation sites ({\it e.g.} between $1$ cm and $10^{2}$ cm). Bearing in
mind the uncertainty in this, we here take a conservative view and
consider a quark drop of initial dimensions $R_{s,0}=10^{5}$ fm, much
below the above range. Considering such a small quark drop implies
restricting our analysis to the very final stages of the transition, but
it is then that the self-similarity is expected to break down and a change
in the hydrodynamical evolution is expected to occur.

	As initial conditions for the time evolution with the full
hydrodynamical equations we therefore use the general form of the
self-similar solutions, which is determined once the degree of
supercooling in either one of the two phases has been established. It is
worth noticing that the supercooling does not need to be extremely small
as is the case during bubble percolation and coalescence. In fact, during
the very final stages of the transition considered here, the quark volume
fraction in the Universe has become very small and the confinement
processes are no longer able to supply the energy necessary to maintain
the increased hadron volume fraction at essentially $T_c$ against the
cooling produced by the expansion of the Universe \cite{fma88}. All of the
models which we present here refer to a quark drop having initial
temperature ${\hat T_q} = T_q/T_c = 0.998$, surrounded by a hadron plasma
at initial temperature ${\hat T_h} = T_h/T_c = 0.990$. A degree of
supercooling of $1\%$ in the hadron phase is, in our view, reasonable and
allows numerical simulations to be performed within acceptable time costs.
Moreover, it should be noted that results obtained with a smaller degree
of supercooling ({\it e.g.} down to $0.1\%$) show only minimal overall
differences for $e$, $\rho$ and $w_0$ (always below a few percent). The
situation is different if the degree of supercooling is chosen to be {\it
larger}. In this case, which probably has no cosmological relevance, the
hydrodynamical evolution can be rather different and would follow the
lines discussed in \cite{rmp95}.

	As mentioned in Section IV, an important feature of the present
simulations is the transition between total coupling of the radiation and
standard fluids and their effective decoupling. While in nature this
process would take place in a rapid but gradual way, the start of it is
necessarily discontinuous when simulated by means of a numerical
calculation on a grid. For this reason it has been necessary to introduce
a free parameter $R_d$, referred to as the ``decoupling radius'', fixing
the drop radius at which the change is made from one regime to the other.
For drop radii \hbox{$R_s > R_d$} the two fluids are considered as totally
coupled and moving as a single fluid. The phase interface is taken to have
a width related to the m.f.p. of the radiation fluid particles and the
characteristics of the radiation fluid are taken to coincide with the ones
of the standard fluids. In practice the coupling is treated by adding the
number of degrees of freedom of the radiation fluid particles to the
number of degrees of freedom in the two phases of the strongly interacting
matter and by setting to zero the contribution of the source functions
$s_0$ and $s_1$ and the energy flux $w_1$. Also, the jump in $w_0$ at the
interface is then calculated in terms of that for $e$. Conversely, for
drop radii $R_s < R_d$ the two fluids are considered as not being totally
coupled and the calculation of the radiation fluid variables adjacent to
the interface is made using the radiation characteristics which are now
distinct from those of the standard fluids. At this stage the radiation
fluid evolves separately from the standard ones and long range energy and
momentum transfer can start to take place.

	It is worth pointing out that while in the above procedure the
decoupling between the two fluids {\it starts} in a discontinuous manner,
the decoupling in itself is {\it gradual} and is governed by the radiation
hydrodynamic equations. The abrupt switch is certainly an approximation
but, as discussed in next Section, it is a rather good one and numerical
results show that the hydrodynamical evolution quickly recovers from the
perturbation introduced by the sudden decoupling.

\vfill\eject
\bigskip
\sezione{VI}{Numerical results}
\subsez{A}{The standard parameters}

	This Section is devoted to the presentation of the results
obtained from the numerical integration of the hydrodynamical
equations for the radiation and the standard fluids. We first present
results for a standard set of the parameters of the problem and will
discuss later the changes introduced when these parameters are allowed
to vary. We here consider an isolated quark drop of initial radius
\hbox{$R_{s,0}=10^5$} fm, surrounded by a hadron plasma at temperature
\hbox{${\hat T_h}= 0.99$}, and to which is associated a phase
interface with surface tension parameter \hbox{$\sigma_0 =
\sigma/T_c^3 = 1$}. (We assume \hbox{$T_c=150$ MeV}). Moreover, we
consider the phase interface as a perfect black-body ({\it i.e.}
$\alpha_1=1$) and the non conservative scattering contribution in the
first source function as maximal ({\it i.e.} $\alpha_2=1$). The
decoupling radius is related to the m.f.p. of the radiation
fluid particles and we here set \hbox{$R_d=\lambda=10^4$} fm.

	Figures 2 and 3 show the time evolution of the radial component of
the Eulerian four-velocity $u$ and the energy density $e$ of the standard
fluids. The phase interface is represented by the vertical discontinuity,
with the quark phase always being to the left of it and with the different
curves referring to different stages during the contraction. The
decoupling is allowed to start at a drop radius of $10^4$ fm; as can be
seen from the graphs, the solution is not particularly perturbed by the
new conditions and quickly returns to a regular behaviour.

\bigskip
\bigskip
\bigskip
\centerline{\vbox{\hsize 12.0truecm\baselineskip=12pt\noindent\tenrm
\textfont1=\tenmit
Figure 2. Time evolution of $u$, the radial component of the fluid
four-velocity in the Eulerian frame. The quark phase is to the left of
the vertical discontinuity. The decoupling between the radiation fluid
and the standard fluids is allowed to start at $R_s=10^4$ fm} }
\bigskip
\bigskip
\bigskip
\centerline{\vbox{\hsize 12.0truecm\baselineskip=12pt\noindent\tenrm
\textfont1=\tenmit
Figure 3. Time evolution of the profile of the energy density $e$ in the
standard fluid.} }
\bigskip
\bigskip
\bigskip

	These graphs are quite similar to the ones presented in
\cite{rmp95} even though in this case the self-similar solution during the
contraction is more weakly preserved after the decoupling. When the drop
reaches dimensions comparable with the intrinsic length scale
\hbox{$\sigma/w_q \approx 10^2$} fm, the self-similar behaviour is
irreversibly lost and the evaporation then proceeds through the
accelerated stages already observed in \cite{rmp95}. This is related to
the contribution of the surface tension which has become overwhelming and
produces a compression within the quark phase with a consequent
temperature increase. Despite the reduced dimensions of the drop surface,
the increased temperature jump between the two phases of the strongly
interacting matter is able to preserve a considerable hydrodynamical flux
away from the surface (the outward velocity is increased), thus allowing
for an increasingly rapid evaporation which ends with the complete
disappearance of the drop. Note that the treatment of the phase interface
as a discontinuity surface is no longer correct when the drop radius
reaches 1 fm and so our results for the smallest radii should only be
treated as indicative.

	Figures 4 and 5 show the time evolution of the radiation energy
density $w_0$ and of the radiation energy flux $w_1$. Before the
decoupling starts, $w_0$ obviously follows the self-similar evolution of
the energy density of the standard fluids and the energy and momentum
transfer between the two types of fluid is so efficient that they can be
considered as in local thermodynamic equilibrium, giving ($w_1=0$). The
situation changes when the drop becomes smaller then $10^4$ fm. At this
stage the decoupling starts and this has the effect of smearing out the
step in the radiation energy density which was present before. Now the
radiative transfer is able to carry away the energy stored within the
radiation fluid in the quark phase.

	As can be seen from the small diagrams in Figures 4 and 5, which
show the evolution of $w_0$ and $w_1$ immediately after the decoupling has
started, this process is quite rapid and before the drop radius has
decreased by one order of magnitude, the radiation energy density profile
has flattened out, equalizing with the value at infinity. The energy flux
$w_1$ deviates from zero and becomes positive as soon as the decoupling
starts and then progressively decays as the step in the radiation energy
density is smeared out. This process is somewhat similar to the rapid
release of the radiative energy contained within an optically thick, hot
but non-emitting gas sphere which suddenly starts to become optically thin
and is allowed to emit.

\bigskip
\bigskip
\bigskip
\centerline{\vbox{\hsize 12.0truecm\baselineskip=12pt\noindent\tenrm
\textfont1=\tenmit
Figure 4. Time evolution of the radiation energy density $w_0$. The
dashed curves are the result of dominant Doppler effects at the
very end of the drop evaporation. The curves in the small diagram show
the rapid evolution of energy density immediately after the decoupling
has started.} }
\bigskip
\bigskip
\bigskip
\centerline{\vbox{\hsize 12.0truecm\baselineskip=12pt\noindent\tenrm
\textfont1=\tenmit
Figure 5. Time evolution of radiation energy flux $w_1$. The curves in
the small diagram show the rapid increase of the energy flux
immediately after the decoupling has started.} }
\bigskip
\bigskip
\bigskip

	The dashed curves in Figure 4 correspond to the very final stages
of the drop evaporation ({\it i.e.} for drop dimensions of the order of a
few fm). The increase in the radiation energy density which is seen there
is related to the motion of the Lagrangian observers with respect to an
essentially uniform radiation field and therefore has a pure Doppler
nature (it can be shown that under these circumstances \hbox{$w_0 \simeq
(1+4u^2/3)(w_0)_{_N}$} and \hbox{$w_1 \simeq -(4 \G u/3)(w_0)_{_N}$},
\cite{rm94}). Note that Doppler contributions are always present after the
decoupling and are more evident in the energy flux, where they enter at
the first order in $u$ and are responsible for the increasing negative
flux observed for drop radii smaller than $10^3$ fm.

	Some of the most interesting effects produced by the decoupling
between the radiation fluid and the standard fluids regard the evolution
of the compression factor $\rho$. As discussed in \cite{rmp95}, a key
property of the self-similar solution is that of preserving the values of
the compression factor in the two phases of the strongly interacting
matter. This reflects a perfect balance between the competing effects of
the compression which would tend to be produced by the reduction in size
of the quark drop and the evaporation processes which extract matter from
it. As pointed out in \cite{rmp95}, an increase in the compression within
the quark phase is possible only when the self-similar solution is broken
and this can occur either when the long range energy and momentum transfer
takes place or, later on, when the drop radius becomes comparable with the
intrinsic length scale related to the surface tension. If the decoupling
between the radiation fluid and the standard fluids is neglected, (as in
\cite{rmp95}), the compression produced is purely hydrodynamical and this
takes place only during the very final stages of the drop evaporation. In
that case, the relative increase of $\rho^+$, (the compression factor
immediately ahead of the phase interface), at the end of the contraction
of a standard quark drop with $\sigma_0=1$ and initial ${\hat T_h}=0.99$,
was computed to be of the order of $40\%$.

	The situation changes dramatically if the radiative transfer
between the standard fluids and the radiation fluid is consistently taken
into account. Figure 6 shows the time evolution of the compression factor
in both phases of the strongly interacting matter. With the magnified
scale it is not possible to see the initial values of the compression
factors which are \hbox{$\rho_h=0.253$} for the hadron phase and
\hbox{$\rho_q=1.0$} for the quark phase, (our reference value). It is
evident that as soon as the decoupling is allowed to take place at $10^4$
fm, the compression within the quark phase starts to increase
progressively and, at the end of evaporation, it has reached values which
are more than two orders of magnitude larger (an increase of
\hbox{$ \sim 5\times 10^4 \; \%$ !)}. The small diagram in Figure 6 traces the
values of the compression just ahead of the phase interface ($\rho^+$) and
just behind it ($\rho^-$).

\bigskip
\bigskip
\bigskip
\centerline{\vbox{\hsize 12.0truecm\baselineskip=12pt\noindent\tenrm
\textfont1=\tenmit
Figure 6. Time evolution of standard fluid compression factor $\rho$.
The curves in the small diagram represent the values of the
compression factor immediately ahead of the phase interface
($\rho^+$) and immediately behind it ($\rho^-$). }}
\bigskip
\bigskip
\bigskip

	The explanation for this striking behaviour is related to the fact
that the long range radiative transfer is able to extract energy from within
the quark phase without extracting the strongly interacting matter. As a
consequence, the relation between the compression factor and the pressure
(and hence between the compression factor and the temperature) is altered
and the evaporation evolves in a radically non-adiabatic manner. The main
effect produced by the radiative transfer is then that of reducing the
specific entropy of the quark-gluon plasma, so that it is possible to
enhance the quark compression without significantly changing its
temperature.

	It is interesting that the growth in the quark compression factor
continues to occur also {\it after} the radiation energy density in the
quark phase has been levelled down to the value in the hadron phase and
the outward energy flux from the quark phase has become very small ({\it
i.e.} even for $R_s < 10^3$ fm). This is due to the fact that when the
energy density of the radiation fluid within the quark phase has reached
the same value as in the hadron phase, there is a local temperature
difference between the radiation fluid and the quark-gluon plasma which
drives a very small but finite energy flux from the quark plasma into the
radiation fluid, where it is then redistributed very efficiently. In this
way the process of entropy extraction from the quark phase is able to
operate even when the outward radiation energy flux from the quark phase
is very small.

	A limit to this mechanism is, of course, introduced by the
intrinsic dimensions of the drop and by the length scale for the
interactions of the particles of the radiation fluid. If the drop is
too small, it becomes effectively transparent to the radiation
particles and the entropy extraction is no longer efficient; at this
stage the decoupling between the two fluids can be considered to be
{\it complete}.  For the typical quark drop under consideration here,
this happens at about $10^2$ fm, where the increase in the compression
factor temporarily slows down (see the small diagram in Figure 6). At
this stage the solution would become self-similar again, but for the
fact that the quark drop is now small enough for the surface tension
to take over and dominate the final stages of the evaporation,
producing the last compression enhancement. In the next Section it
will be shown that it is possible to recover the self-similar solution
again after the total decoupling has taken place if a suitable choice
of the decoupling radius and of the m.f.p. $\lambda$ is made (Figure
11).

	A special comment should be made concerning a result which we
consider to be particularly important. As discussed before, our
treatment of the long range energy and momentum transfer between the
radiation fluid and the standard fluids leads to an increase in the
compression factor of the {\it quark} phase by about two orders of
magnitude. It should be kept in mind, however, that this peak value is
limited to a very small volume (of the order of 1 fm$^3$) and that it
would be dispersed by the rarefaction wave following the complete
disappearance of the drop (see \cite{rmp95} for a description of the
rarefaction wave and Section VII for further discussions). As a
consequence, if a relic inhomogeneity from the transition is to be
investigated, this should rather concern the compression seen in the {\it
hadron} phase before the disappearance of the drop.

	Figure 7 shows the final profile of the compression factor $\rho$
computed when the quark drop has reached a radius of 1 fm. It is
interesting to note that besides the large peak in the quark phase, the
compression factor has been increased also in the hadron phase, where it
appears as a plateau of comparatively smaller magnitude. However, if one
selects a vertical scale with greater resolution and normalizes the values
of the compression factor to the background hadron compression (see the
small diagram of Figure 7), it is clear that the plateau does indeed have
a specific profile, with a maximum about two orders of magnitude larger
than the background value. More important, the hadron compression extends
over a much larger length scale, which coincides with the interaction
length scale of the radiation fluid particles. Figures 6 and 7 could give
a misleading impression as they seem to show that the most important
effect is the compression increase in the quark phase whereas, in fact,
the relative compression increase in the hadron phase is also substantial
and is more significant in that it extends over a volume which is twelve
orders of magnitude larger.

\bigskip
\bigskip
\bigskip
\centerline{\vbox{\hsize 12.0truecm\baselineskip=12pt\noindent\tenrm
\textfont1=\tenmit
Figure 7. Final profile of the compression factor $\rho$; the
computation has been stopped when the quark drop has a radius of 1
fm. The small diagram shows, with a different vertical resolution, the
same profile after it has been normalized to the value of the hadron
compression at infinity. }}
\bigskip
\bigskip
\bigskip

	The compression increase in the hadron phase is not produced
directly by the radiation, but rather results from the fact that
``over-compressed'' quark fluid elements (with decreased specific entropy)
give rise to ``over-compressed'' hadron fluid elements after they have
undergone the phase transformation in accordance with the junction
condition (\ref{ejc}). (Note that the entropy {\it increase} which
naturally occurs across the phase interface is much smaller that the {\it
decrease} introduced by the radiation fluid, so that fluid elements in
the hadron phase near the drop have smaller specific entropy than those
far from it). A key point to stress is that the over-compressed hadron
plasma is in pressure balance (and therefore in temperature balance) with
the surrounding hadron medium. This is a consequence of the decrease of
specific entropy which took place while the fluid elements concerned were
still inside the drop. At the end of the transition a spherical region of
over-compressed hadron plasma is left behind which is in equilibrium with
the surrounding medium. This is the region where a baryon number
concentration could be produced and this would then only be dispersed by
neutron diffusion on the time scale relevant for that. The consequences of
this result for the production of baryon number inhomogeneities at the end
of the transition will be discussed in Section VII.

\bigskip
\subsez{B}{The parameter space}

	In this Section we discuss the changes introduced for the drop
evaporation by variation of the set of the parameters within the allowed
parameter space. We start by commenting on the hydrodynamical evolution of
a quark drop for which the coefficient $\alpha_1$, which relates the
hydrodynamical flux to the thermal flux in the hadron phase, is not equal
to unity as in the case of a perfect \hbox{black-body} surface. In
general, reducing $\alpha_1$ has the effect of decreasing the
``transparency'' of the drop surface to the phase transformation and
therefore of slowing down the drop evaporation and favoring the long range
energy and momentum transfer away from the quark phase.

\bigskip
\bigskip
\bigskip
\centerline{\vbox{\hsize
12.0truecm\baselineskip=12pt\noindent\tenrm \textfont1=\tenmit
Figure 8. Compression factors immediately ahead of and behind the phase
interface when the radius of the quark drop has decreased to 1 fm, as a
function of the adjustable coefficient $\alpha_1$. The dashed curves
extrapolate the numerical results to very small values of $\alpha_1$, for
which computations are not possible. } }
\bigskip
\bigskip
\bigskip

	Figure 8 shows the variation, as a function of $\alpha_1$, of the
compression factors immediately ahead of and behind the phase interface
when the radius of the quark drop has decreased to 1 fm (the other
parameters are left unchanged from the values discussed in the previous
Section). While the solid curves fit points obtained by single
computations, the dashed curves are an extrapolation of these to values of
$\alpha_1$ for which the computations would have been exceedingly time
consuming (the computational time tends to infinity as $\alpha_1$ tends to
zero). It is interesting to notice that the formation of high compressions
in the quark and hadron phases is a general feature and that the
relative increase of the compression factors in both phases can easily be
of six or seven orders of magnitude, thus giving a stronger cosmological
relevance to this process.

	Let us now consider the changes brought about by variation of
the non-conservative scattering coefficient $\alpha_2$ in the energy
source moment (\ref{s0c}). As mentioned in Section II, rough estimates
indicate that $\alpha_2 \approx 1$ in the present cosmological scenario,
but it is nevertheless interesting to consider situations for smaller
values of $\alpha_2$. It is obvious that a larger non-conservative
scattering coefficient will enhance the efficiency of the radiative
transfer processes and, in turn, the formation of compressed regions of
the strongly interacting fluids.

\bigskip
\bigskip
\bigskip
\centerline{\vbox{\hsize 12.0truecm\baselineskip=12pt\noindent\tenrm
\textfont1=\tenmit
Figure 9. Compression factor immediately ahead of the phase interface
for computations with different values of the adjustable coefficient
$\alpha_2$. The small diagram shows the equivalent curves for the
compression factor immediately behind the phase interface. } }
\bigskip
\bigskip
\bigskip

	As shown in Figure 9, where results of computations performed with
five different values of $\alpha_2$ are presented, the hydrodynamical
evolution is not qualitatively changed and although a value of
$\alpha_2=1$ maximizes the compression, a relative compression increase
(at the end of the drop evaporation) of about two orders of magnitude is
present also in the total absence of the scattering contribution.

	All of the results discussed so far are from simulations in which
the decoupling between the radiation fluid and the standard fluids was
allowed to start at a ``decoupling radius'' $R_d$ equal to the m.f.p.
$\lambda$ of the strongly interacting particles. While such a choice is
suggested by elementary considerations, there is no reason to exclude
slightly smaller or larger values of $R_d$ and it was interesting to
consider the changes introduced for a decoupling started at $R_d/ \lambda
\not= 1$.

\bigskip
\bigskip
\bigskip
\centerline{\vbox{\hsize 12.0truecm\baselineskip=12pt\noindent\tenrm
\textfont1=\tenmit
Figure 10. Compression factor immediately ahead of the phase interface
when the radius of the quark drop has decreased to 1 fm, as a function of
the decoupling radius $R_d$. The vertical axis is normalized to the value
of $\rho^+$ obtained for $R_d=\lambda$. The solid line fits points
obtained by single numerical simulations and the small diagram magnifies the
results for small values of the decoupling radius. } }
\bigskip
\bigskip
\bigskip

	Figure 10 collects the results of this investigation presenting
the values of the final quark compression (at $R_s=1$ fm), for different
values of the decoupling radius. The value of $\rho^+$ presented in the
diagram is normalized to the value obtained for $R_d=\lambda$; $R_d=1.2
\;\lambda$ is the largest value for which a satisfactory numerical
solution could be obtained, but $R_d=\lambda$ leads to a more regular
behaviour of the hydrodynamical quantities and so was used for the
standard run presented in the previous Section.

	The interpretation of Figure 10 is straightforward: making the
decoupling at smaller values of the drop radius has the effect of
reducing the time interval during which the long range energy and
momentum transfer away from the quark phase takes place. As a
consequence, the specific entropy in the quark phase is changed less,
leading to a smaller final compression. If the value of $R_d/\lambda$ is
taken to be very small, the hydrodynamical behaviour tends to the one
observed when the decoupling is totally neglected and ultimately
coincides with the solution obtained in \cite{rmp95} when
$R_d/\lambda=0$. This is a satisfying result and shows that the
numerical modelling has an overall physical consistency.

	Another example of this coherence appears when a self-similar
solution can be recovered after decoupling between the radiation
fluid and the standard fluids is complete. This can be produced if
$\lambda$ is artificially increased so as to be much larger than the
length scale associated with the surface tension, thus separating the
two possible regimes during which a compression can be produced.

	Figure 11 shows the profiles of the compression factors
immediately ahead of and behind the phase interface for values of
$\lambda=R_d$ ranging between $10^4$ fm (the physically realistic value)
and $10^7$ fm. (In all simulations the quark drop has initial dimensions
$R_{s,0}=10 \lambda$.) It is evident that with the standard set of
parameters, (shown with the continuous line), the two different
compression growth stages join together and that self-similar evolution
(represented by a constant compression factor state) cannot set in. The
situation is rather different for the (unrealistic) choice of
$R_d=\lambda=10^7$ fm. In this case it is possible to distinguish clearly
between the initial compression growth (produced by the relativistic
radiative transfer), and the final compression enhancement (a consequence
of the accelerated evaporation driven by the surface tension) which in all
of the simulations takes place for $R_s \ltord 10^2$ fm. The evolution
between the two stages clearly follows a self-similar solution and this
seems to be a further example of the widespread occurrence of the
self-similar solutions for an isolated contracting spherically symmetric
system.

\bigskip
\bigskip
\bigskip
\centerline{\vbox{\hsize 12.0truecm\baselineskip=12pt\noindent\tenrm
\textfont1=\tenmit
Figure 11. Compression factors immediately ahead of and behind the
phase interface. Different curves refer to different values of the
m.f.p. of the radiation fluid particles (expressed in fm) and show
that if a large enough value is chosen, a self-similar evolution is
reached. All curves are drawn for $R_d=\lambda$ and initial quark
dimensions one order of magnitude larger than $\lambda$. } }
\bigskip
\bigskip
\bigskip

	A final comment in this Section should be made concerning the role
played by the neutrinos in the process of long range energy and momentum
transfer away from the quark phase. As mentioned in Section II, neutrinos
have been neglected in the present calculation because of the much larger
length scale at which they interact ($\lambda_{\nu} \approx 10^{13}$ fm).
Nevertheless, on this scale they can be considered as particles of a
generalized radiation fluid and could operate a radiative transfer process
similar to the one discussed so far for the electromagnetically
interacting particles and produce a compressed hadron medium at the end of
their decoupling.

	In order to investigate the amplitude of this compression, we
have performed a computation in which we simulate the decoupling between
a radiation fluid composed only of neutrinos, and a standard fluid
composed of strongly and electromagnetically interacting particles. It
should be noted that this is a rather speculative investigation since it
assumes the existence of isolated, spherical quark regions of dimensions
at least comparable with $\lambda_{\nu}$, and it is not clear whether
the disconnection of quark regions happens at a scale large enough for
this to occur. However, bearing this reservation in mind, results of our
calculations for the effects of neutrino decoupling on the compression
profiles are presented in Figure 12.

	It is evident that entropy extraction by means of neutrinos is
less effective than for the case of the electromagnetically interacting
particles and this is the result of the different combination of the
number of the degrees of freedom in the two cases (for neutrinos
$g_r=5.25$). Nevertheless, the decoupling produces a non negligible
compression in both phases, giving a compression in the hadron plasma
which is about five times greater than the background one.

\bigskip
\bigskip
\bigskip
\centerline{\vbox{\hsize 12.0truecm\baselineskip=12pt\noindent\tenrm
\textfont1=\tenmit
Figure 12. Compression factors immediately ahead of and behind the
phase interface. Here $R_d=\lambda_{\nu}=10^{13}$ fm.} }
\bigskip
\bigskip
\bigskip

	Reduction of $\alpha_1$ would lead to further amplification of the
compression in the same way as already discussed for the decoupling of the
electromagnetically interacting particles. This result is also relevant for
considerations of the baryon number density profile which is left behind
by the quark-hadron transition and will be further discussed in the next
Section.

\sezione{VII}{Cosmological implications}

	We here briefly discuss some of the consequences that the results
presented in the previous Sections can have in a cosmological context. A
more detailed analysis of these features will be presented in a
forthcoming paper \cite{mpr95}.

	A first question concerns the relation between the compression
factor $\rho$ in the two phases of the strongly interacting matter and the
baryon number density which has a more direct physical relevance.
Certainly, most of the astrophysical consequences which have been
discussed in relation with a first order quark-hadron phase transition are
connected with the production of baryon number inhomogeneities which could
survive until later epochs. Baryon number density has a natural tendency
to be discontinuous across the phase interface since baryon number is
carried by almost massless quarks in the high temperature phase, while in
the low temperature phase it is carried by heavy nucleons whose number
density is strongly suppressed. In the limit of chemical equilibrium
across the front, the baryon chemical potentials are equal for both phases
of the strongly interacting matter, and the net baryon flow across the
phase interface vanishes. If the evolution of the transition is isothermal
with both phases at $\approx T_c$, the ratio of baryon number densities
$k=(n_{_B}^q/n_{_B}^h)$ can be easily computed after having specified the
critical temperature \cite{fma88}, with $k$ being $\gtord 10$ for $T_c
\ltord 150$ MeV. The values obtained are slightly larger if the finite
volume of the hadrons is not neglected \cite{bmm85}.

	This baryon number segregation can be further enhanced when the
chemical equilibrium is broken (this could either occur because the
interface velocity is much larger than the mean baryon diffusion velocity
or because the bubbles have dimensions larger than the typical baryon
diffusion scale length in a Hubble time $R_{_B}^{diff} \sim 10^{10}$ fm).
As a result of the breaking of chemical equilibrium, baryon number could
accumulate on the quark side of the front. Depending on the intensity of
the net baryon number flow into the hadron region and on the efficiency of
diffusion in smearing out accumulations of baryon number, the baryon
density contrast can be magnified between 2 and 6 orders of magnitude
\cite{afm87,jfmk94}. In addition to its relevance for cosmological
nucleosynthesis, this mechanism has also been considered in connection
with the possibility of it giving rise to primordial magnetic fields which
could serve as seeds for the production of the present intergalactic
and interstellar magnetic fields \cite{co94}. We here note that seed
magnetic fields might also be produced at the very end of the transition
when the quark drops evaporate rapidly and the radiative transfer favours
baryon number segregation and hence charge separation \cite{mpr95}.

	The evolution of the baryon number contrast has been investigated
by Kurki-Suonio \cite{ks88} who considered several scenarios for the
creation of the final baryon number density profile arising from various
combinations of the intrinsic scale lengths of the problem: {\it i.e.} the
baryon diffusion length, the mean separation of nucleation sites, and the
typical dimension of hadron bubbles at coalescence. The last two length
scales in particular, are still uncertain today and it has not yet been
possible to clarify further the situation described in \cite{ks88}. One of
the scenarios considered by Kurki-Suonio was concerned with baryon number
concentration produced by long range radiative transfer and, within this
context, the results presented in the previous Sections can be used to
provide an updated view of this. In the simplest picture where baryon
number is taken to be strictly advected along with the hydrodynamical
flow, the baryon number density is directly proportional to the
compression factor $\rho$ and so Figures 6--12 can be considered as
representing also the baryon number density. However, departures from this
proportionality can be caused both by diffusion of baryon number (which
can operate when the typical length scale for variations in $\rho$ is
smaller than the relevant diffusion length scale or comparable with it),
or by suppression of baryon number flow across the phase interface (which
would lead to a build up of baryon number in the quark phase).  It is
expected that a filter mechanism would operate at the phase interface
accumulating baryon number there and augmenting the concentration produced
by the specific entropy extraction via the radiation fluid particles
\cite{fma88}. As a consequence, the results presented for the compression
in the quark phase represent a lower limit to the possible enhancement of
baryon number density in the high temperature phase.

	As mentioned in Section VI, the scale length of the
inhomogeneities produced by this mechanism is given by the m.f.p. for the
radiative fluid particles and for the electromagnetically interacting
particles this is much smaller than the minimum inhomogeneity scale length
that can affect nucleosynthesis. An underlying large amplitude baryon
number contrast would have to be produced during the intermediate stages
of the transition in order for the baryon number segregation produced to
be on a large enough scale to be able to survive and be relevant for
nucleosynthesis. This contrast might be achieved at the time of hadron
bubble coalescence, or possibly during the decoupling of neutrinos from
the standard fluids as discussed at the end of Section VI.

	A final interesting issue to be investigated is the hydrodynamical
evolution of the compression enhancements after the quark phase has been
totally converted into the hadron one. Concentrating on a single quark
drop, it is easy see that a rarefaction wave (possibly fronted by a
spherical shock) should appear when the drop disappears \cite{rmp95,
kks86}. At this stage, the source of the outward flow from the quark phase
ceases to exist and the flow profile should progressively deform as it
moves out into the compressible hadron medium. As a result of this
deformation, a shock front could be produced and this would then be
followed by a region where the medium compressed by the shock is rarefied
again to an equilibrium value. Given the energies and velocities, the it
might well be that no shock appears or that there is only a rather weak
shock which would damp rapidly. Independently of the fine details of the
mechanism, the overall effect will be that of redistributing the excess
energy and compression which was within the very small region of the
disappearing quark drop. A numerical computation would be required to
provide a full description of this process \cite{mpr95}, but it is
possible to make a rough estimate of the eventual degree of dilution of
the compressed matter after the disappearance of the phase interface.

%\vfill\eject

	For this purpose, consider the sum of the enthalpy contained
within a quark drop of radius $R_s \sim 1/T_c \sim 1$ fm

\begin{equation}
W\approx (e+p)_q \times {4\over3} \pi T_c^{-3} =
{4\over3} \pi^2
g_q T_c^4  \times {4\over3} \pi T_c^{-3}
\approx 136.0 \ {\rm fm^{-1}},
\end{equation}
\noindent
and the surface energy

\begin{equation}
\Sigma \approx \sigma_0 T_c^3 \times 4 \pi T_c^{-2} \approx 25.1 \
{\rm fm^{-1}} .
\end{equation}
\noindent
Taking this to be converted into enthalpy of hadronic matter
when the drop disappears, we then have an overdense region with
enthalpy density

\begin{equation}
(e+p)^{\prime} = {(W + \Sigma) \over {4/3 \pi T_c^{-3}}}
\approx 307.5 \ {\rm fm^{-4}} ,
\end{equation}
\noindent
which will subsequently expand and come into equilibrium with the
surrounding medium in which

\begin{equation} (e+p)_h = 4 g_h \pi^2 T_c^4 / 90 \approx 21.1 \ {\rm
fm^{-4}}.  \end{equation} \noindent If we make the assumption that the
specific entropy of the material which was within the drop after the
disappearance of the phase interface remains essentially unchanged, we
then have \hbox{$ (e+p) \propto \rho^{4/3}$}. In this case, compression
dilution can be estimated to be

\begin{equation}
{\rho_{{fin}} \over{\rho^{\prime}}} =
\left[ {(e+p)_h \over {(e+p)^{\prime} } } \right]^{3/4}
\approx {1\over{7.5}}
\end{equation}
\noindent
where $\rho_{_{fin}}$ is the final compression factor of the fluid
element and $\rho^{\prime}$ is its compression factor when the phase
interface disappeared.

\sezione{VIII}{Conclusion }

	In this paper we have discussed the relativistic hydrodynamics of
the very final stages of the cosmological quark-hadron phase transition.
In particular, we have studied the evaporation of a single isolated
spherical quark drop including the effects of long range energy and
momentum transfer by means of electromagnetically interacting particles.
This transfer takes place when the quark drop reaches dimensions which are
comparable with the mean free paths of these particles and can lead to a
significant modification of the hydrodynamical evolution (see \cite{rmp95}
for comparison). For this study, a set of Lagrangian hydrodynamical
equations for describing the evolution of the strongly interacting fluids
has been coupled to an equivalent set of equations describing the
hydrodynamics of the fluid of electromagnetically interacting particles. A
numerical code has been constructed for integrating the complete set of
equations and results from the computations have been presented.

	The evolution of the quark drop starts by following the
self-similar solution which characterizes an isolated spherical
evaporating configuration and this behaviour is then broken when
decoupling of the radiation fluid from the standard fluids takes place. A
particular consequence of the long range energy and momentum transfer is
the establishment of an entropy flux away from the quark phase carried by
the \hbox{long-m.f.p.} particles of the radiation fluid. This acts so as
to increase the compression of both phases of the standard fluid in the
vicinity on the drop, producing overall relative increases of two orders
of magnitude or more. Thus, even in the absence of suppression mechanisms
operating at the phase interface, contrasts in the baryon number density
of several orders of magnitude are natural products of a first order
quark-hadron phase transition. The hydrodynamical properties of this
process are completely general and similar results have been obtained when
exploring the whole parameter space of the problem. In particular, it has
also been shown that larger compressions (up to seven orders of magnitude)
can easily be achieved if the transparency of the phase interface to the
hydrodynamical flux is decreased.

	These computations, which are the first simulations of quark drop
evaporation in the presence of radiative transfer, provide useful
quantitative information about the final stages of the transition which
can be used in studies of the evolution of baryon number inhomogeneities
\cite{mpr95}. Density peaks of baryon number could be associated with the
production of primordial magnetic fields, generated by the charge
separation across the phase interface, and could possibly affect
nucleosynthesis if produced over a large enough length scale.

%\acknowledgments
{\bigskip\medskip\bigbf{Acknowledgments} \bigskip}

	We gratefully acknowledge helpful discussions with Ornella
Pantano. Financial support for this research has been provided by the
Italian Ministero dell'Universit\`a e della Ricerca Scientifica e
Tecnologica.

\vfill\eject

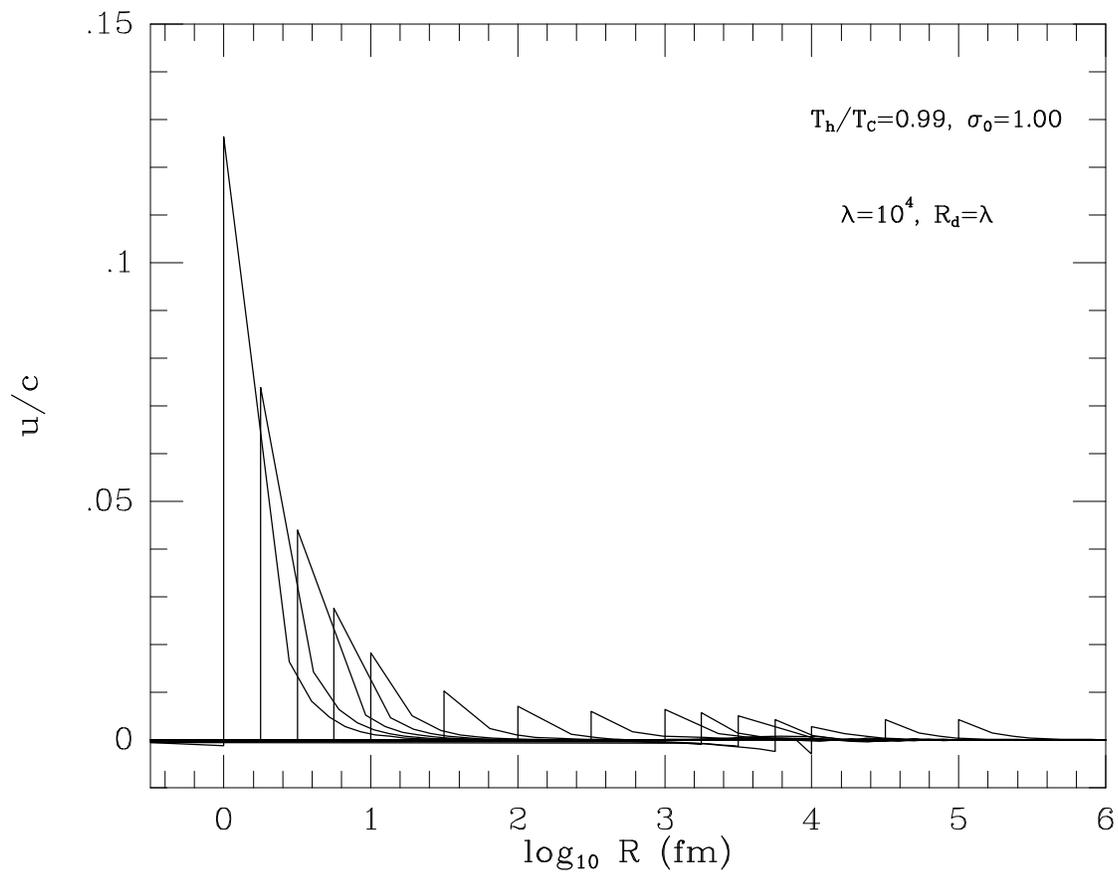
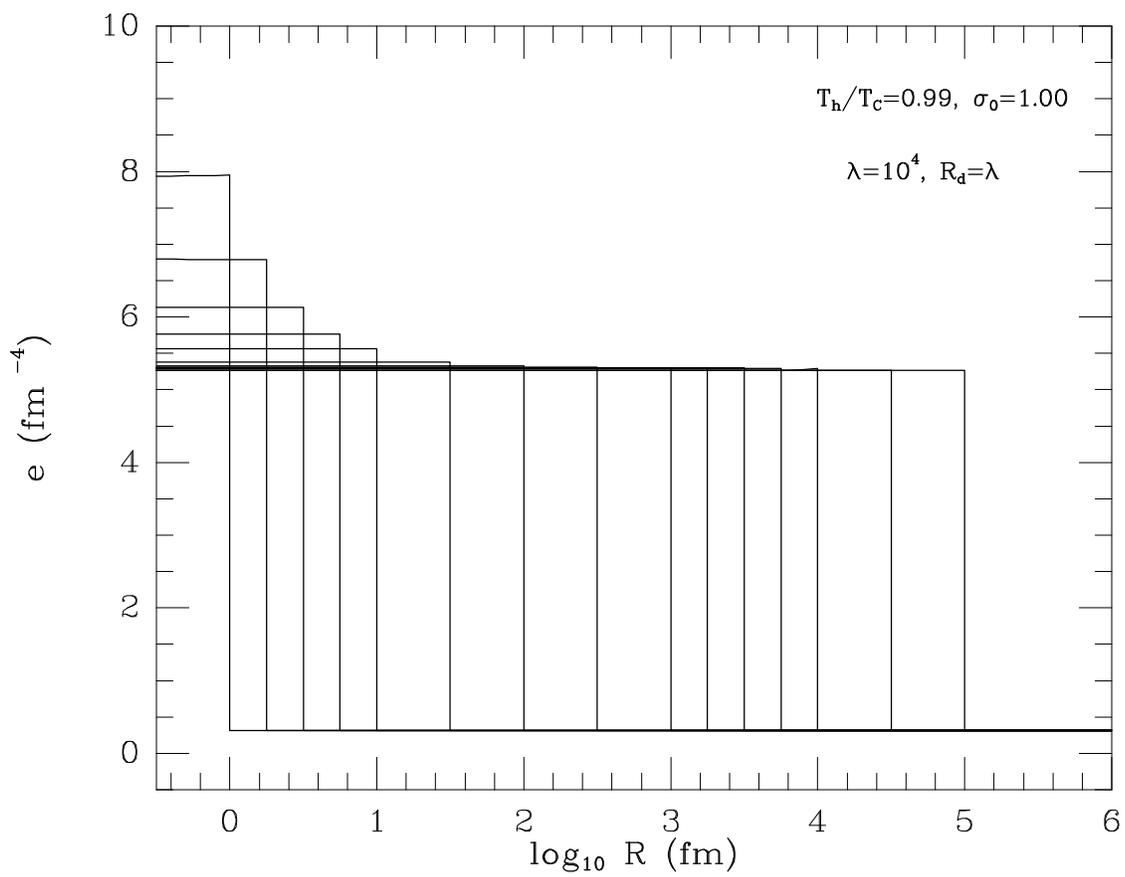
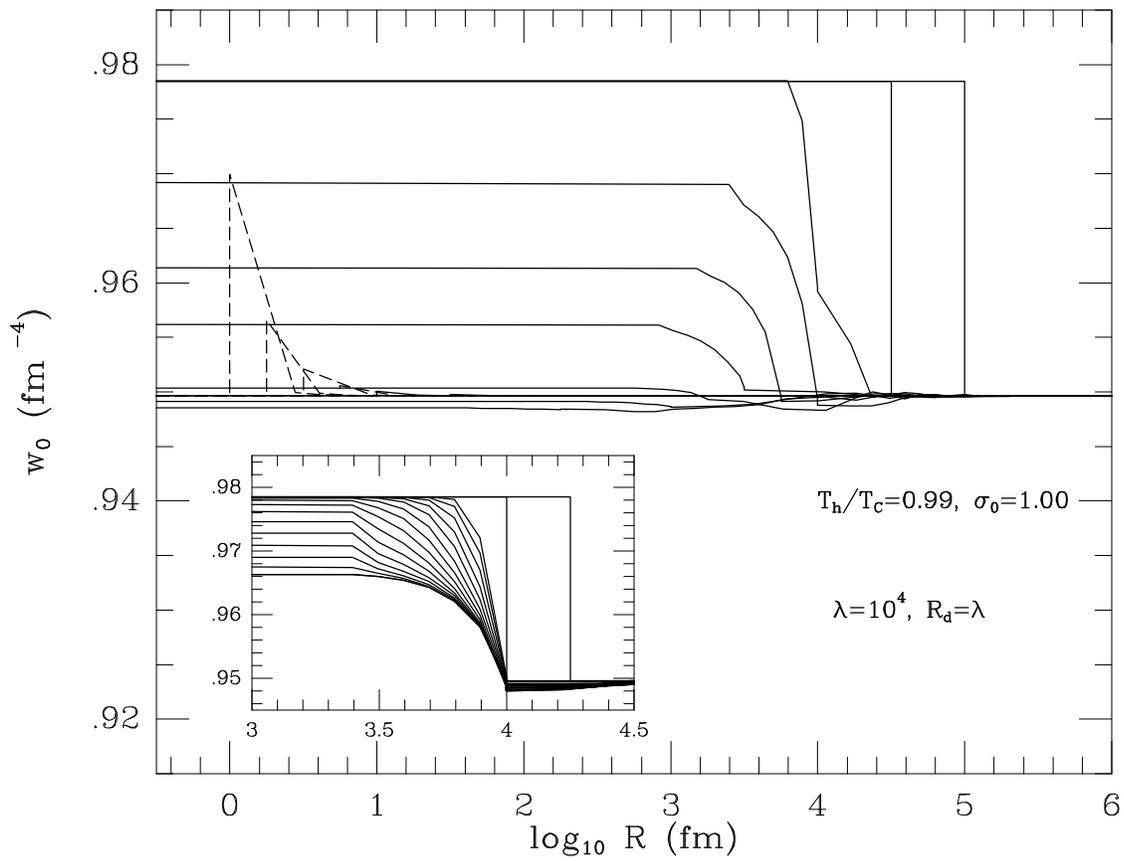
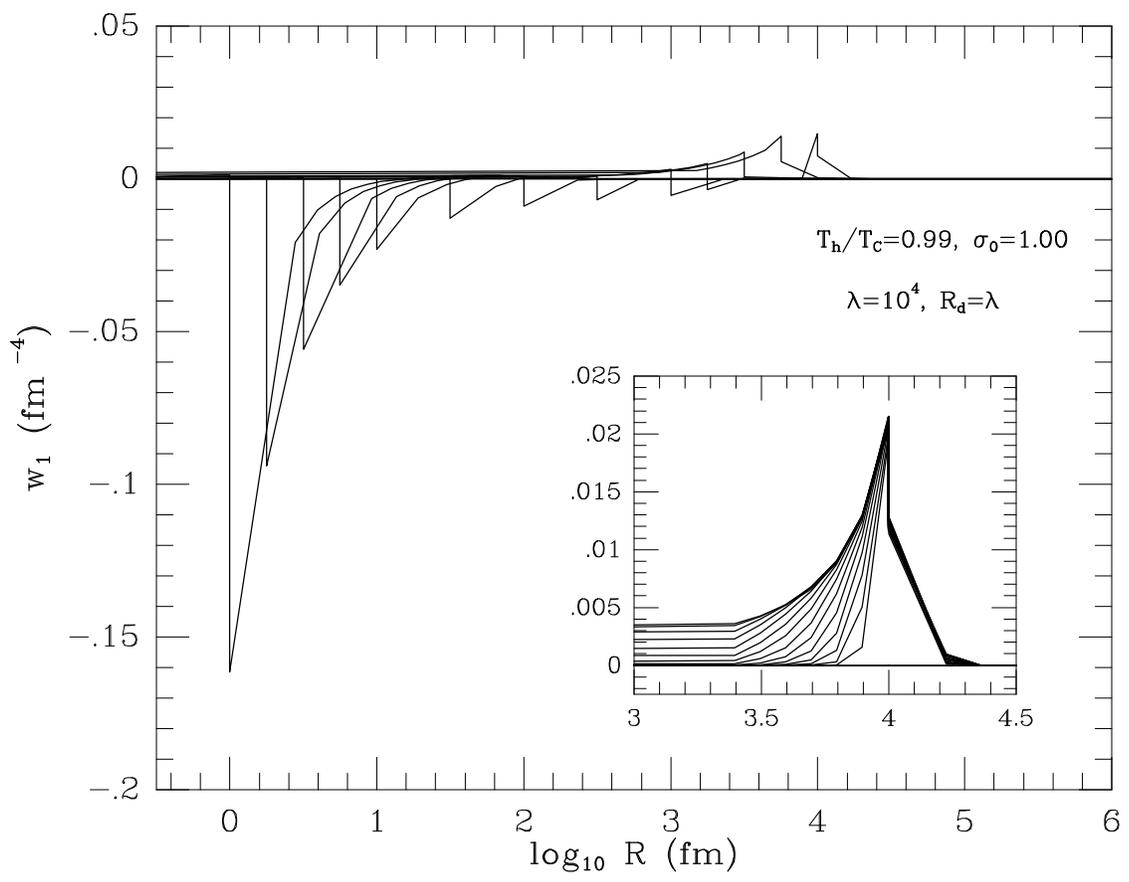
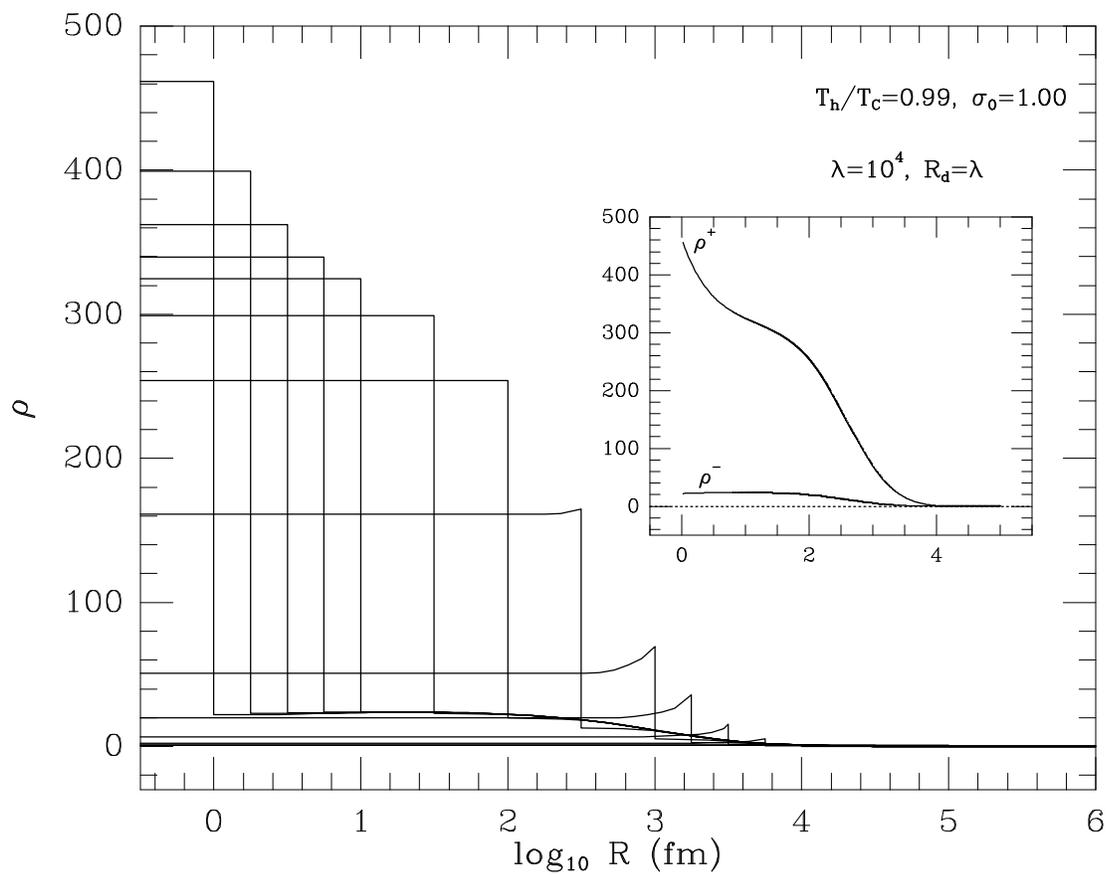
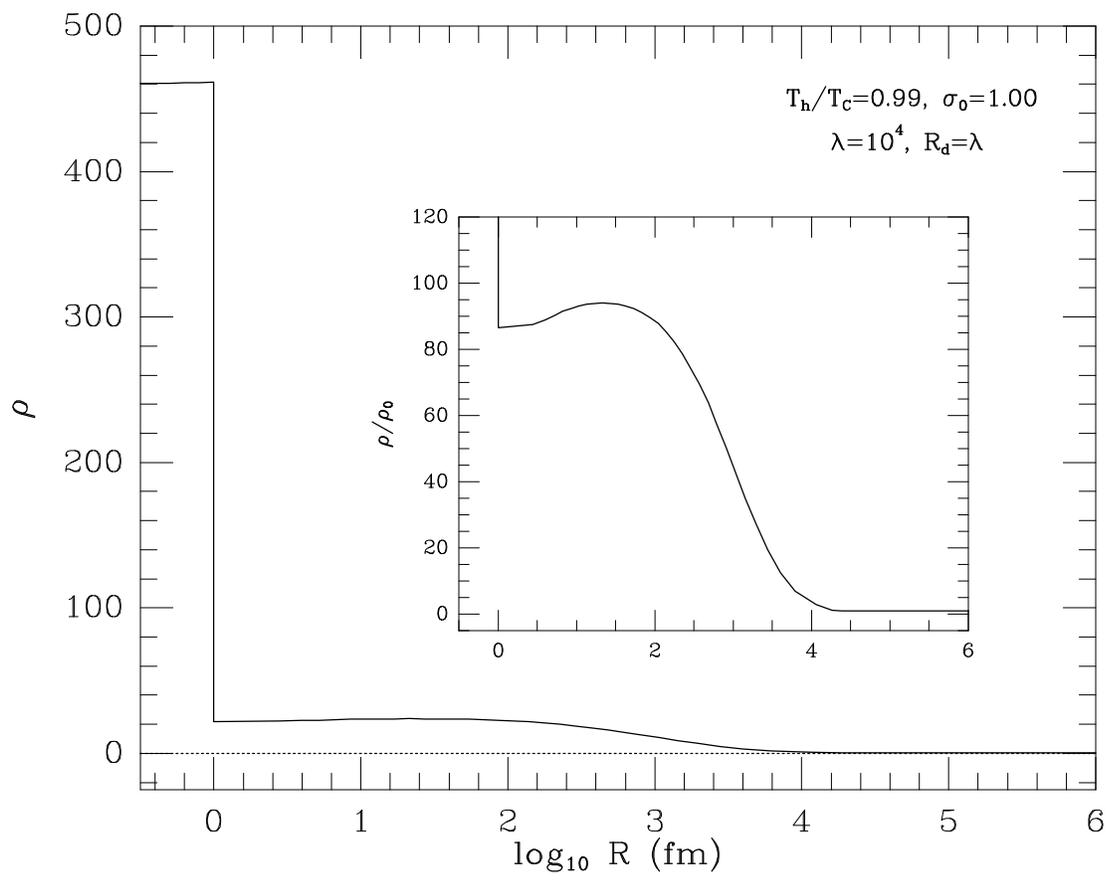
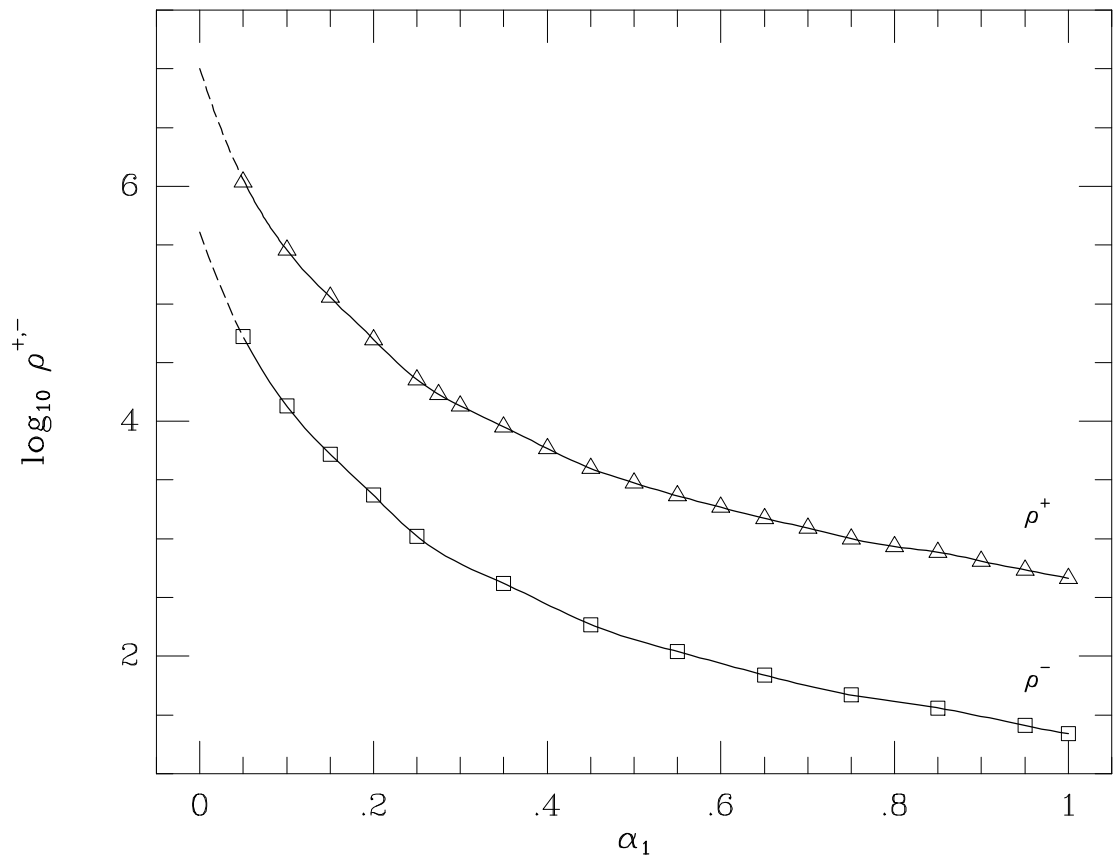
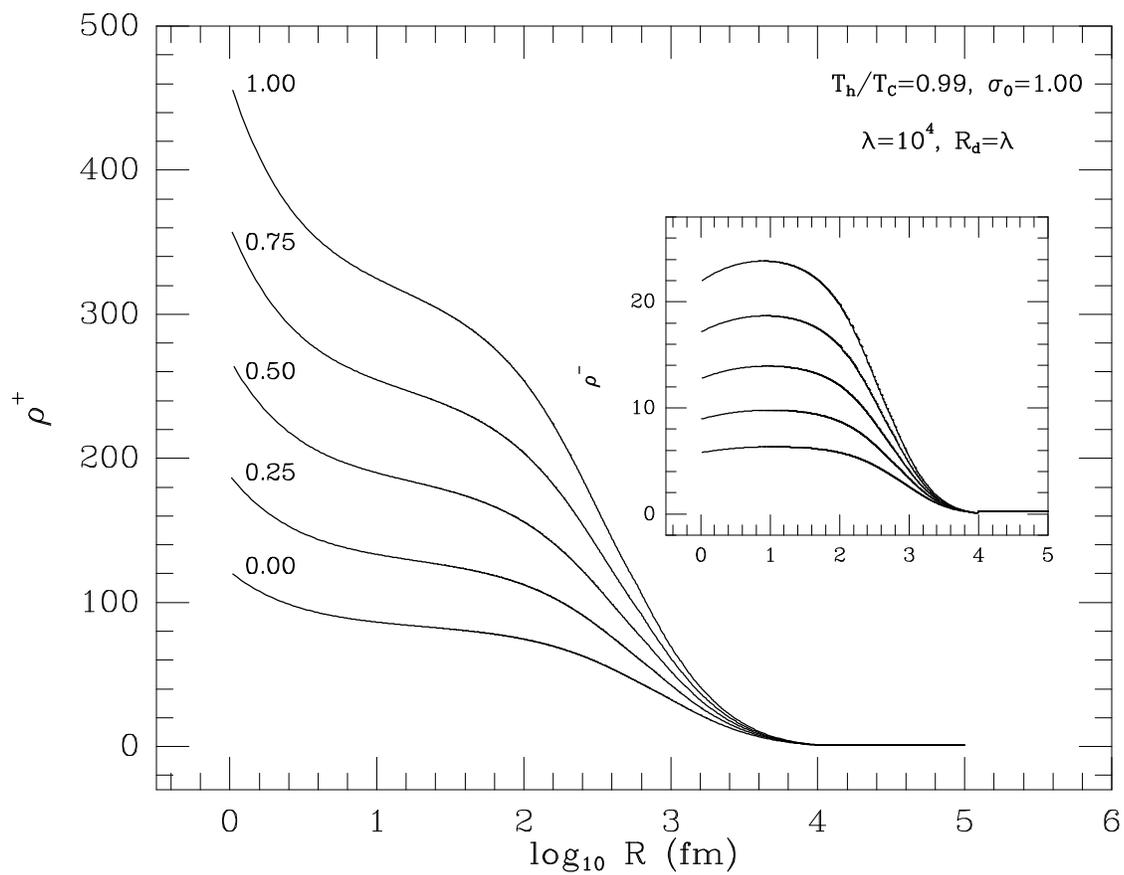
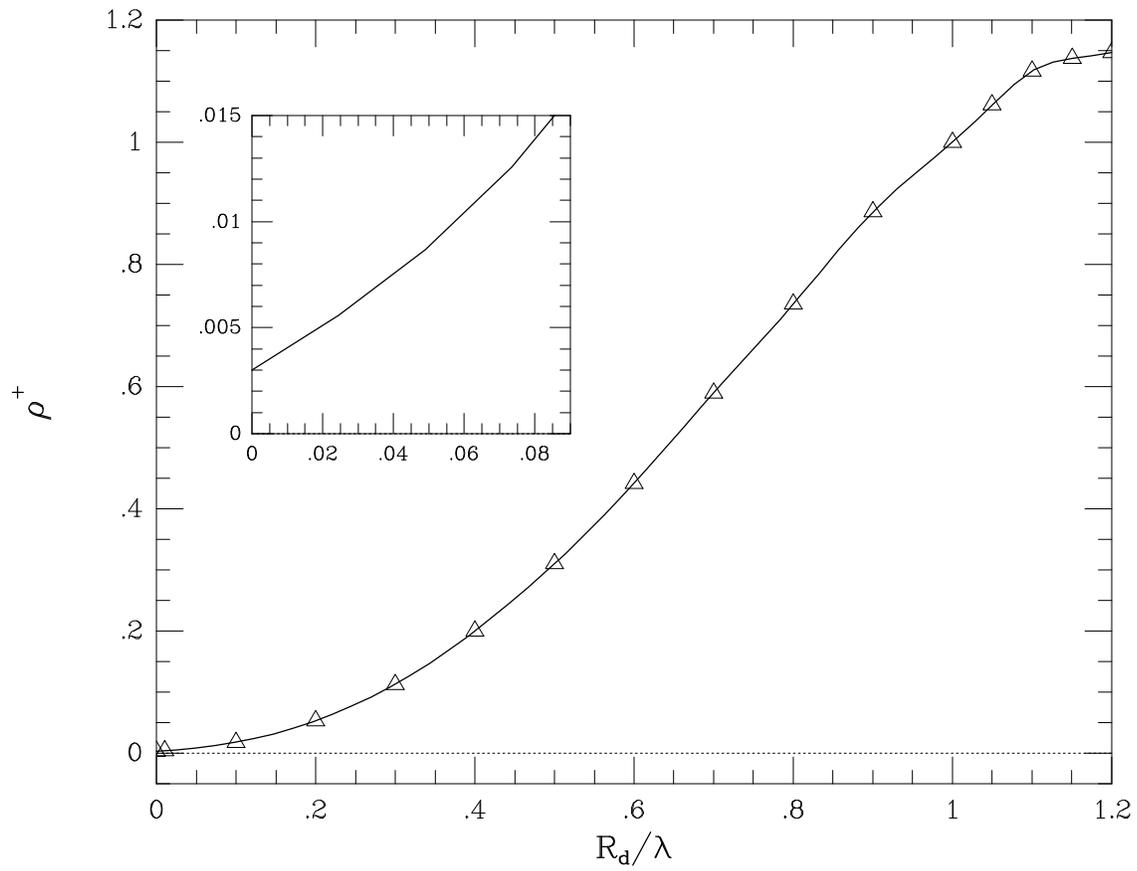
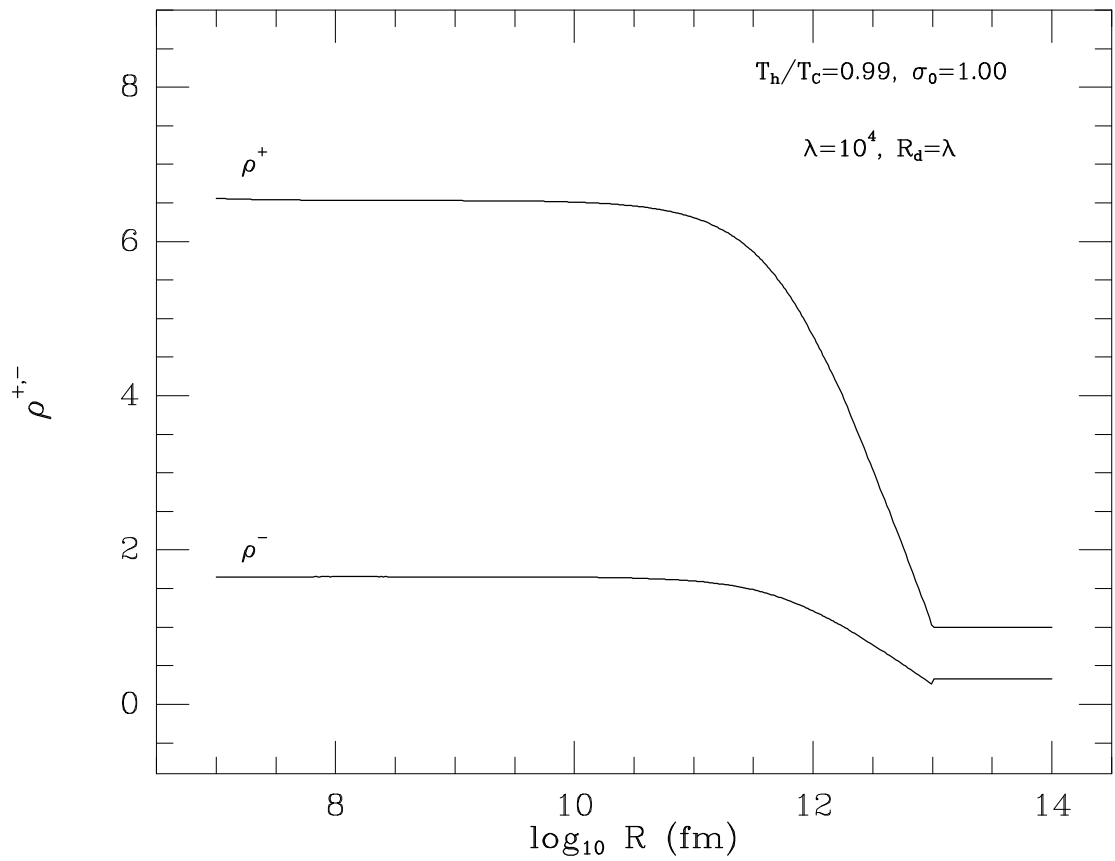
\end{document}